# A Cryptographic Treatment of the Wiretap Channel

Mihir Bellare[1] Stefano Tessaro[2] Alexander Vardy[3]

April 2010

**Abstract**

The wiretap channel is a setting where one aims to provide information-theoretic privacy of communicated data based solely on the assumption that the channel from sender to adversary is "noisier" than the channel from sender to receiver. It has been the subject of decades of work in the information and coding (I&C) community. This paper bridges the gap between this body of work and modern cryptography with contributions along two fronts, namely *metrics* (definitions) of security, and *schemes*. We explain that the metric currently in use is weak and insufficient to guarantee security of applications and propose two replacements. One, that we call mis-security, is a mutual-information based metric in the I&C style. The other, semantic security, adapts to this setting a cryptographic metric that, in the cryptography community, has been vetted by decades of evaluation and endorsed as the target for standards and implementations. We show that they are equivalent (any scheme secure under one is secure under the other), thereby connecting two fundamentally different ways of defining security and providing a strong, unified and well-founded target for designs. Moving on to schemes, results from the wiretap community are mostly non-constructive, proving the existence of schemes without necessarily yielding ones that are explicit, let alone efficient, and only meeting their weak notion of security. We apply cryptographic methods based on extractors to produce explicit, polynomial-time and even practical encryption schemes that meet our new and stronger security target.

[1] Department of Computer Science & Engineering, University of California San Diego, 9500 Gilman Drive, La Jolla, California 92093, USA. Email: `mihir@cs.ucsd.edu`. URL: `http://www.cs.ucsd.edu/users/mihir`. Supported in part by NSF grants CNS-0904380, CNS-1116800 and CCF-0915675 and DARPA contract HR011-09-C-0129.

[2] Department of Computer Science & Engineering, University of California San Diego, 9500 Gilman Drive, La Jolla, California 92093, USA. Email: `stessaro@cs.ucsd.edu`. URL: `http://www.cs.ucsd.edu/users/stessaro`. Supported in part by Calit2 and NSF grant CNS-0716790.

[3] Department of Electrical and Computer Engineering, University of California San Diego, 9500 Gilman Drive, La Jolla, California 92093, USA. Email: `avardy@ucsd.edu`. URL: `http://www.ece.ucsd.edu/~avardy/`. Supported in part by DARPA contract HR011-09-C-0129.

# Contents

# 1 Introduction

This paper aims to bridge the gap between two communities. The first, within information and coding (I&C), is the wiretap channel community, and the second is the modern cryptographic community.

The wiretap channel is a setting where one aims to communicate data with information-theoretic security under the sole assumption that the channel from sender to adversary is "noisier" than the channel from sender to receiver. Introduced by Wyner, Csiszár and Körner in the late seventies [45, 14], it has developed in the I&C community over the last 30 years divorced from the parallel development of Modern Cryptography. Yet the questions, centering as they are on data security, are at heart cryptographic.

The first element of the gap is definitions. We explain that the security definition in current use, that we call mis-r (mutual-information security for random messages) is weak and insufficient to provide security of applications. We suggest strong, new definitions. One, that we call mis (mutual-information security), is an extension of mis-r and thus rooted in the I&C tradition and intuition. Another, semantic security, adapts a definition of [19, 2] that, vetted by decades of cryptographic research and targeted by standards and schemes deployed in practice, is the cryptographic gold standard. We prove the two equivalent, thereby connecting two fundamentally different ways of defining privacy and providing a new, strong and well-founded target for constructions.

The second element of the gap is techniques. The I&C community tends to view security as an add-on to error-correction, starting from error-correcting codes (ECCs) and tweaking the designs to confer security. Their results are mostly non-constructive, proving existence of schemes whose algorithms may not be efficient. Meanwhile, complexity-theorists and cryptographers have developed tools such as extractors which would seem eminently useful in this domain, even more so when one aims to achieve the new and more stringent definitions of privacy that we define. With this starting point, we take the opposite approach, making error-correction an add-on to security. We start from cryptographic tools to provide the security, error-correction being done later for no purpose other than correcting errors. We obtain constructive results which yield practical schemes to achieve all the security goals we define. Let us now look at all this in more detail.

The wiretap model. The setting is depicted in Figure 1. The sender applies to her message $\mathsf{M}$ a randomized encryption function $\mathcal{E}\colon \{0,1\}^m \to \{0,1\}^c$ to get what we call the *sender-ciphertext* $\mathsf{X} \leftarrow\!\$\, \mathcal{E}(\mathsf{M})$.[1] This is transmitted to the receiver over the receiver channel $\mathsf{ChR}$ so that the latter gets a *receiver ciphertext* $\mathsf{Y} \leftarrow\!\$\, \mathsf{ChR}(\mathsf{X})$ which it decrypts via algorithm $\mathcal{D}$ to recover the message. The adversary's wiretap is modeled as another channel $\mathsf{ChA}$ and it accordingly gets an *adversary ciphertext* $\mathsf{Z} \leftarrow\!\$\, \mathsf{ChA}(\mathsf{X})$ from which it tries to glean whatever it can about the message.

A *channel* is a randomized function specified by a transition probability matrix $W$ where $W[x,y]$ is the probability that input $x$ results in output $y$. Here $x, y$ are strings. Thus, for example, we regard the Binary Symmetric Channel $\mathsf{BSC}_p$ with crossover probability $p \leq 1/2$ as taking a binary string $x$ of any length and returning the string $y$ of the same length formed by flipping each bit of $x$ independently with probability $p$. For concreteness and simplicity of exposition we will often phrase discussions in the setting where $\mathsf{ChR}, \mathsf{ChA}$ are BSCs with crossover probabilities $p_R, p_A \leq 1/2$ respectively, but our results apply in much greater generality. In this case the assumption that $\mathsf{ChA}$ is "noisier" than $\mathsf{ChR}$ corresponds to the assumption that $p_R < p_A$. This is the only assumption made: the adversary is computationally unbounded, and the scheme is keyless, meaning sender and receiver are not assumed to a priori share any information not known to the adversary.

Previous work. The setting now has a literature, in the I&C community, encompassing hundreds of papers. (See the survey [27] or the book [5].) Schemes must satisfy two conditions, namely *decoding* and *security*. The *decoding* condition asks that the scheme provide error-correction over the receiver channel, namely $\lim_{m\to\infty} \Pr[\mathcal{D}(\mathsf{ChR}(\mathcal{E}(\mathsf{M}))) \neq \mathsf{M}] = 0$. The original security condition of [45] was that $\lim_{k\to\infty} \mathbf{I}(\mathsf{M}; \mathsf{ChA}(\mathcal{E}(\mathsf{M}))/m = 0$ where $k$ is an underlying security parameter of which $m, c$ are functions, the message random variable $\mathsf{M}$ is uniformly distributed over $\{0,1\}^m$ and $\mathbf{I}(\mathsf{M}; \mathsf{Z}) = \mathbf{H}(\mathsf{M}) - \mathbf{H}(\mathsf{M} \mid \mathsf{Z})$ is the mutual information. This was critiqued by [32, 36] who put forth the stronger security condition now in use, namely that $\lim_{k\to\infty} \mathbf{I}(\mathsf{M}; \mathsf{ChA}(\mathcal{E}(\mathsf{M})) = 0$. (The random variable $\mathsf{M}$ continues to be uniformly distributed over $\{0,1\}^m$.) The literature seeks to minimize the *rate* $\mathbf{Rate}(\mathcal{E}) = m/c$ of the scheme.

[1] (The notation $y \leftarrow\!\$\, A(x)$ means that we run randomized function $A$ on input $x$ and denote the output by $y$.

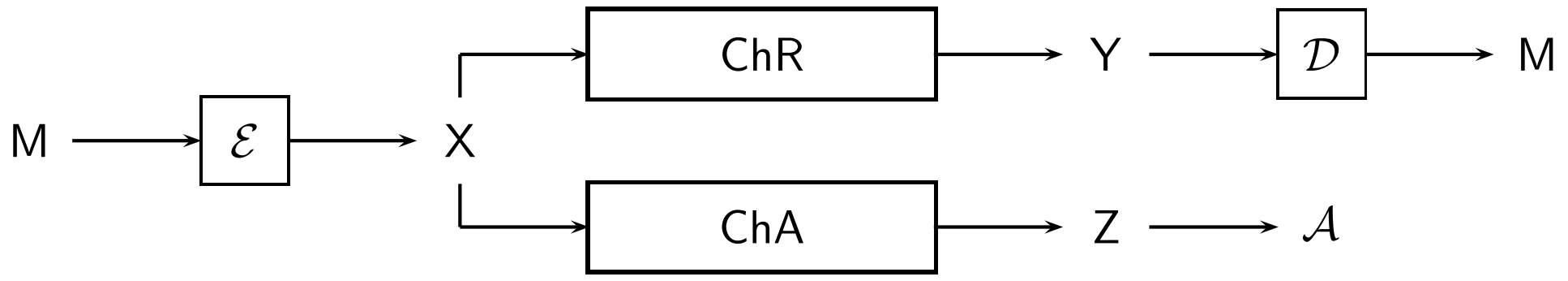

Figure 1: **Wiretap channel model.** See text for explanations.

Shannon's seminal result [41] says that if we ignore security and merely consider achieving the decoding condition then the maximum achievable rate $\mathbf{Rate}(\mathcal{E})$ is the receiver channel capacity, which in the BSC case is $1 - h_2(p_R)$ where $h_2$ is the binary entropy function $h_2(p) = -p \lg(p) - (1-p)\lg(1-p)$. He gave non-constructive proofs of existence of schemes meeting capacity.

Coming in with this background and the added security condition, it was natural for the wiretap community to follow Shannon's lead and begin by asking what is the maximum achievable rate, now subject to both the security and decoding conditions. This optimal rate is called the secrecy capacity and, in the case of BSCs, equals the difference $(1 - h_2(p_R)) - (1 - h_2(p_A)) = h_2(p_A) - h_2(p_R)$ in capacities of the receiver and adversary channels. Non-constructive proofs of the existence of schemes with this optimal rate were given in [45, 14, 6]. A lot of work has followed aiming to establish similar results for other channels. Little attention has been given to finding explicit schemes with efficient encoding and decoding.

Context. Practical interest in the wiretap setting is escalating. Its proponents note two striking benefits over conventional cryptography: (1) no computational assumptions, and (2) no keys and hence no key distribution. Item (1) is attractive to governments who are concerned with long-term security and worried about quantum computing. Item (2) is attractive in a world where vulnerable, low-power devices are proliferating and key-distribution and key-management are unsurmountable obstacles to security. The practical challenge is to realize a secrecy capacity, meaning ensure by physical means that the adversary channel is noisier than the receiver one. The degradation with distance of radio communication signal quality is the basis of several approaches being investigated for wireless settings. Government-sponsored Ziva Corporation [46] is using optical techniques to build a receiver channel in such a way that wiretapping results in a degraded channel. A program called Physical Layer Security aimed at practical realization of the wiretap channel is the subject of books [5] and conferences [22]. All this activity means that schemes are being sought for implementation. If so, we need privacy definitions that yield security in applications, and we need constructive results yielding practical schemes achieving privacy under these definitions. This is what we aim to supply.

Definitions. A security *metric* xs associates to encryption function $\mathcal{E}\colon \{0,1\}^m \to \{0,1\}^c$ and adversary channel $\mathsf{ChA}$ a number $\mathbf{Adv}^{\mathrm{xs}}(\mathcal{E}; \mathsf{ChA})$ that measures the maximum "advantage" of an adversary in breaking the scheme under metric xs. For example, the metric underlying the current, above-mentioned security condition is $\mathbf{Adv}^{\text{mis-r}}(\mathcal{E}; \mathsf{ChA}) = \mathbf{I}(\mathsf{M}; \mathsf{ChA}(\mathcal{E}(\mathsf{M})))$ where $\mathsf{M}$ is uniformly distributed over $\{0,1\}^m$. We call this the mis-r (mutual-information security for random messages) metric because messages are assumed to be random. From the cryptographic perspective, this is extraordinarily weak, for we know that real messages are not random. (They may be files, votes or any type of structured data, often with low entropy. Contrary to a view in the I&C community, compression does *not* render data random, as can be seen from the case of votes, where the message space

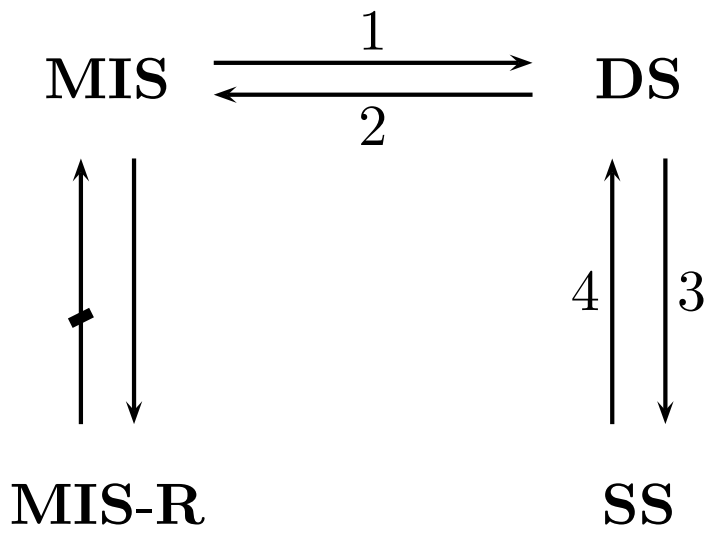

| | | |
|---|---|---|
| 1 | $\epsilon_{\mathrm{ds}} \le \sqrt{2\epsilon_{\mathrm{mis}}}$ | Theorem 4.5 |
| 2 | $\epsilon_{\mathrm{mis}} \le 2\epsilon_{\mathrm{ds}} \lg \frac{2^c}{\epsilon_{\mathrm{ds}}}$ | Theorem 4.9 |
| 3 | $\epsilon_{\mathrm{ss}} \le \epsilon_{\mathrm{ds}}$ | Theorem 4.1 |
| 4 | $\epsilon_{\mathrm{ds}} \le 2\epsilon_{\mathrm{ss}}$ | Theorem 4.1 |

Figure 2: **Relations between notions.** An arrow $\mathbf{A} \to \mathbf{B}$ is an implication, meaning every scheme that is **A**-secure is also **B**-secure, while a barred arrow $\mathbf{A} \not\to \mathbf{B}$ is a separation, meaning that there is a **A**-secure scheme that is not **B**-secure. If $\mathcal{E}\colon\ \{0,1\}^m \to \{0,1\}^c$ is the encryption function and $\mathsf{ChA}$ the adversary channel, we let $\epsilon_{\mathrm{xs}} = \mathbf{Adv}^{\mathrm{xs}}(\mathcal{E};\mathsf{ChA})$. The table then shows the quantitative bounds underlying the annotated implications.

---

has very low entropy.) This leads us to suggest a stronger metric that we call *mutual-information security*, defined via $\mathbf{Adv}^{\mathrm{mis}}(\mathcal{E};\mathsf{ChA}) = \max_{\mathsf{M}} \mathbf{I}(\mathsf{M};\mathsf{ChA}(\mathcal{E}(\mathsf{M})))$ where the maximum is over all random variables $\mathsf{M}$ over $\{0,1\}^m$, regardless of their distribution.

At this point, we have a legitimate metric, but how does it capture privacy? The intuition is that it is measuring the difference in the number of bits required to encode the message before and after seeing the ciphertext. This intuition is alien to cryptographers, whose metrics are based on much more direct and usage-driven privacy requirements. Cryptographers understand since [19] that encryption must hide all partial information about the message, meaning the adversary should have little advantage in computing a function of the message given the ciphertext. (Examples of partial information about a message include its first bit or even the XOR of the first and second bits.) The mis-r and mis metrics ask for nothing like this and are based on entirely different intuition. We extend Goldwasser and Micali's semantic security [19] definition to the wiretap setting, defining

$$\mathbf{Adv}^{\mathrm{ss}}(\mathcal{E};\mathsf{ChA}) \;=\; \max_{f,\mathsf{M}} \left( \max_{\mathcal{A}} \Pr[\mathcal{A}(\mathsf{ChA}(\mathcal{E}(\mathsf{M}))) = f(\mathsf{M})] - \max_{\mathcal{S}} \Pr[\mathcal{S}(m) = f(\mathsf{M})] \right) .$$

Within the parentheses is the maximum probability that an adversary $\mathcal{A}$, given the adversary ciphertext, can compute the result of function $f$ on the message, minus the maximum probability that a simulator $\mathcal{S}$ can do the same given only the length of the message. We also define a distinguishing security (ds) metric as an analog of indistinguishability [19] via

$$\mathbf{Adv}^{\mathrm{ds}}(\mathcal{E};\mathsf{ChA}) \;=\; \max_{\mathcal{A},M_0,M_1} 2\Pr[\mathcal{A}(M_0, M_1, \mathsf{ChA}(\mathcal{E}(M_{\mathsf{b}}))) = \mathsf{b}] - 1$$

where challenge bit $\mathsf{b}$ is uniformly distributed over $\{0,1\}$ and the maximum is over all $m$-bit messages $M_0, M_1$ and all adversaries $\mathcal{A}$. For any metric xs, we say $\mathcal{E}$ provides XS-security over $\mathsf{ChA}$ if $\lim_{k\to\infty} \mathbf{Adv}^{\mathrm{xs}}(\mathcal{E};\mathsf{ChA}) = 0$.

Relations. The mutual information between message and ciphertext, as measured by mis, is, as noted above, the change in the number of bits needed to encode the message created by seeing the ciphertext. It is not clear why this should measure privacy in the sense of semantic security. Yet we are able to show that mutual-information security and semantic security are equivalent, meaning an encryption scheme is MIS-secure if and only if it is SS-secure. Figure 2 summarizes this and other relations we establish that between them settle all possible relations. The equivalence between SS and DS is the information-theoretic analogue of the corresponding well-known equivalence in the computational setting [19, 2]. As there, however, it brings the important benefit that we can now work with the technically simpler DS. We then show that MIS implies DS by reducing to Pinsker's inequality. We show conversely that DS implies MIS via a general relation between mutual information and statistical distance. As Figure 2 indicates, the asymptotic relations are all underlain by concrete quantitative and polynomial relations between the advantages.

We show that in general MIS-R does not imply MIS, meaning the former is strictly weaker than the latter. We do this by exhibiting an encryption function $\mathcal{E}$ and channel $\mathsf{ChA}$ such that $\mathcal{E}$ is MIS-R-secure relative to $\mathsf{ChA}$ but MIS-insecure relative to $\mathsf{ChA}$. Furthermore we do this for the case that $\mathsf{ChA}$ is a BSC. However in Section 4.6 we show that for certain encryption schemes, namely ones that are separable and message-linear, and for certain channels, MIS-R security does imply MIS-security. This somewhat surprising result has been exploited in [29, 28] to build MIS-secure schemes.

Schemes. We improve over existing schemes in several ways. First, while previous work targeted MIS-R security, we target and achieve the stronger MIS, DS and SS goals. Second, while previous results were largely proving existence of schemes without giving explicit ones, our schemes are not only explicit but have efficient and simple encryption and decryption and may be used in the practical settings that are now emerging in this area. Third, the methods of previous work were ECC-intrusive, taking ECCs and modifying them to add security. Our approach is modular, combining extractors with existing ECCs.

A common misconception is to think that privacy and error-correction may be completely decoupled, meaning one would first build a scheme that is secure when the receiver channel is noiseless and then add an ECC on top to meet the decoding condition with a noisy receiver channel. This does not work because the error-correction helps the adversary by reducing the noise over the adversary channel. The two requirements do need to be considered together. We are still, however, able to provide a modular approach. First designing secure schemes for the case of a noiseless receiver channel, we then put ECCs on top for error-correction, but we are able to state certain simple conditions on the ECCs that suffice for preservation of security, and provide non-intrusive analyses showing that most ECCs have these properties.

Let $\mathcal{H}\colon \{0,1\}^h \times \{0,1\}^u \to \{0,1\}^m$ be a universal hash function. In the case of a noiseless receiver channel, our scheme, given a $m$-bit message $M$, picks at random a $h$-bit string $H$ and $u$-bit string $U$, and sets the sender ciphertext to $U\|H\|\mathcal{H}(H,U)\oplus M$. Using the extraction properties of universal hash functions established by the Leftover Hash Lemma [20] and its generalization [16, 15] we can bound the DS-advantage of this scheme in terms of the probability that the adversary, given the adversary ciphertext, recovers $U$. This probability can in turn be bounded for specific channels like the BSC. In the case the receiver channel has error, we apply an ECC to $U$ to get $U'$, apply an ECC to $H\|\mathcal{H}(H,U)\oplus M$ to get $V'$ and let the sender ciphertext be $U'\|V'$. In this case we bound the DS-advantage of this scheme in terms of what we call the rs-r-advantage of the first ECC, which is the probability that an adversary, given the result of the adversary channel on $U'$, can recover $U$. This probability too can be bounded directly for BSCs. We also provide tools to bound it for other channels. Our approach for the case of a noisy receiver channel is inspired by the concepts of secure sketches and fuzzy extractors [16, 15], translating ideas there to the wiretap setting.

Related work. Mahdavifar and Vardy [28, 29] provide an explicit MIS-R-secure scheme with optimal rate, meaning rate equal to the MIS-R secrecy capacity. But they give no proof that decryption (decoding) is possible for their scheme, even in principle let alone in polynomial time. The central open question in the wiretap channel community was whether there is a polynomial time (this means both encryption and decryption are polynomial time) MIS-R secure scheme with optimal rate. Our schemes achieve MIS-R security in polynomial time but do not have optimal rate, so we do not answer this question. The question has finally been settled by [3], who provide a polynomial-time MIS-R secure scheme with optimal rate.

Mahdavifar and Vardy [28, 29] apply our results from Section 4.6 to conclude that their MIS-R secure scheme also provides MIS (and thus by our results DS and SS) security for certain channels. This shows that the optimal rate for DS-security is the same as for MIS-R security for these channels. The question this raises is whether one can achieve DS-security at this optimal rate in polynomial time. This question too is resolved by [3], whose above-mentioned scheme in fact directly achieves

DS-security.

Appendix A provides a comprehensive survey of the large body of work related to wiretap security, and more broadly, to information-theoretic secure communication in a noisy setup [30].

# 2 Preliminaries

Basic notation and definitions. If $s$ is a binary string then $s[i]$ denotes its $i$-th bit and $|s|$ denotes its length. If $S$ is a set then $|S|$ denotes its size. If $x$ is a real number then $|x|$ denotes its absolute value. If $s_1, \ldots, s_l$ are strings then $s_1\|\cdots\|s_l$ denotes their concatenation. If $s$ is a string and $n$ a non-negative integer then $s^n$ denotes the concatenation of $n$ copies of $s$.

A probability distribution is a function $P$ that associates to each $x$ a probability $P(x) \in [0,1]$. The support $\textsc{supp}(P)$ is the set of all $x$ such that $P(x) > 0$. All probability distributions in this paper are discrete. Associate to random variable $\mathsf{X}$ and event $E$ the probability distributions $P_X, P_{X|E}$ defined for all $x$ by $P_{\mathsf{X}}(x) = \Pr[\mathsf{X} = x]$ and $P_{\mathsf{X}|E}(x) = \Pr[\mathsf{X} = x \mid E]$. We denote by $\lg(\cdot)$ the logarithm in base two, and by $\ln(\cdot)$ the natural logarithm. We adopt standard conventions such as $0 \lg 0 = 0 \lg \infty = 0$ and $\Pr[E_1|E_2] = 0$ when $\Pr[E_2] = 0$. The function $h$: $[0,1] \to [0,1]$ is defined by $h(x) = -x \lg x$. The (Shannon) entropy of a probability distribution $P$ is defined by

$$\mathbf{H}(P) = \sum_x h(P(x)) = -\sum_x P(x) \lg P(x)$$

The (Shannon) entropy of a random variable $\mathsf{X}$ is defined by

$$\mathbf{H}(\mathsf{X}) = \mathbf{H}(P_{\mathsf{X}}) = \sum_x h(P_{\mathsf{X}}(x)) \ .$$

The statistical difference between probability distributions $P, Q$ is defined by

$$\mathbf{SD}(P; Q) = \frac{1}{2} \cdot \sum_x |P(x) - Q(x)| \ .$$

The statistical difference between random variables $\mathsf{X}, \mathsf{Y}$ is defined by

$$\mathbf{SD}(\mathsf{X}_1; \mathsf{X}_2) = \mathbf{SD}(P_{\mathsf{X}_1}; P_{\mathsf{X}_2}) = \frac{1}{2} \cdot \sum_x |\Pr[\mathsf{X}_1 = x] - \Pr[\mathsf{X}_2 = x]| \ .$$

If $\mathsf{X}, \mathsf{Y}$ are random variables the conditional entropy is defined via

$$\mathbf{H}(\mathsf{X} \mid \mathsf{Y}) = \sum_y P_{\mathsf{Y}}(y) \cdot \mathbf{H}(\mathsf{X} \mid \mathsf{Y} = y) \quad \text{where} \quad \mathbf{H}(\mathsf{X} \mid \mathsf{Y} = y) = \sum_x h(P_{\mathsf{X}|\mathsf{Y}=y}(x)) \ .$$

The mutual information between random variables $\mathsf{X}, \mathsf{Y}$ is defined by

$$\mathbf{I}(\mathsf{X}; \mathsf{Y}) = \mathbf{H}(\mathsf{X}) - \mathbf{H}(\mathsf{X} \mid \mathsf{Y}) \ .$$

The guessing-probability $\mathbf{GP}$ and min-entropy $\mathbf{H}_\infty$ of a random variable $\mathsf{X}$ are defined via

$$\mathbf{GP}(\mathsf{X}) \ = \ \max_x \Pr[\mathsf{X} = x] \ = \ 2^{-\mathbf{H}_\infty(\mathsf{X})} \ .$$

The average guessing probability $\mathbf{GP}(\mathsf{X}|\mathsf{Z})$ and average min-entropy $\mathbf{H}_\infty(\mathsf{X}|\mathsf{Z})$ are defined for random variables $\mathsf{X}, \mathsf{Z}$ via

$$\mathbf{GP}(\mathsf{X}|\mathsf{Z}) \ = \ \sum_z \Pr[\mathsf{Z} = z] \cdot \max_x \Pr[\mathsf{X} = x \mid \mathsf{Z} = z] \ = \ 2^{-\mathbf{H}_\infty(\mathsf{X}|\mathsf{Z})} \ .$$

Transforms. We say that $T$ is a transform with domain $D$ and range $R$, written $T$: $D \to R$, if $T(x)$ is a random variable over $R$ for every $x \in D$. Thus, $T$ is fully specified by a sequence $P = \{P_x\}_{x \in D}$ of probability distributions over $R$, where $P_x(y) = \Pr[T(x) = y]$ for all $x \in D$ and $y \in R$. We call $P$ the distribution associated to $T$. This distribution can be specified by a $|D|$ by $|R|$ transition probability matrix $W$ defined by $W[x, y] = P_x(y)$. An adversary too is a transform, and so is a simulator.

Channels. A channel is, again, just a transform. In more conventional communications terminology,

a channel $\mathsf{Ch}\colon D \to R$ has input alphabet $D$ and output alphabet $R$.

If $\mathsf{B}\colon D \to Z$ is a channel and $c \geq 1$ is an integer we define the channel $\mathsf{B}^c\colon \{0,1\}^c \to Z^c$ by $\mathsf{B}^c(X) = \mathsf{B}(X[1])\|\cdots\|\mathsf{B}(X[c])$ for all $X = X[1]\ldots X[c] \in \{0,1\}^c$. The applications of $\mathsf{B}$ are all independent, meaning that if $W$ is the transition probability matrix of $\mathsf{B}$ then the transition probability matrix $W_c$ of $\mathsf{B}^c$ is defined by $W[X,Y] = W[X[1],Y[1]]\cdot\ldots\cdot W[X[c],Y[c]]$ for all $X = X[1]\ldots X[c] \in \{0,1\}^c$ and $Y = Y[1]\ldots Y[c] \in Z^c$. We say that a channel $\mathsf{Ch}$ is binary if it equals $\mathsf{B}^c$ for some channel $\mathsf{B}$ and some $c$, in which case we refer to $\mathsf{B}$ as the base (binary) channel and $\mathsf{Ch}$ as the channel induced by $\mathsf{B}$.

By $\mathsf{BSC}_p\colon \{0,1\} \to \{0,1\}$ we denote the binary symmetric channel with crossover probability $p$ $(0 \leq p \leq 1/2)$. Its transition probability matrix $W$ has $W[x,y] = p$ if $x \neq y$ and $1-p$ otherwise for all $x,y \in \{0,1\}$. The induced channel $\mathsf{BSC}_p^c$ flips each input bit independently with probability $p$.

The receiver and adversary channels of the wiretap setting will have domain $\{0,1\}^c$, where $c$ is the length of the sender ciphertext, and range $\{0,1\}^d$, where the output length $d$ may differ between the two channels. Such channels may be binary, which is the most natural example, but our equivalences between security notions hold for all channels, even ones that are not binary.

If $\mathsf{Ch1}\colon \{0,1\}^{c_1} \to \{0,1\}^{d_1}$ and $\mathsf{Ch2}\colon \{0,1\}^{c_2} \to \{0,1\}^{d_2}$ are channels then $\mathsf{Ch1}\|\mathsf{Ch2}$ denotes the channel $\mathsf{Ch}\colon \{0,1\}^{c_1+c_2} \to \{0,1\}^{d_1+d_2}$ defined by $\mathsf{Ch}(x_1\|x_2) = \mathsf{Ch1}(x_1)\|\mathsf{Ch2}(x_2)$ for all $x_1 \in \{0,1\}^{c_1}$ and $x_2 \in \{0,1\}^{c_2}$. We say that a channel $\mathsf{Ch}\colon \{0,1\}^c \to \{0,1\}^d$ is $(c_1,c_2)$-splittable if there are channels $\mathsf{Ch1}\colon \{0,1\}^{c_1} \to \{0,1\}^{d_1}$ and $\mathsf{Ch2}\colon \{0,1\}^{c_2} \to \{0,1\}^{d_2}$ such that $\mathsf{Ch} = \mathsf{Ch1}\|\mathsf{Ch2}$.

For any integer $s$ we let $\mathsf{Id}_s\colon \{0,1\}^s \to \{0,1\}^s$ denote the identity function defined by $\mathsf{Id}_s(x) = x$ for all $x \in \{0,1\}^s$. This represents a clear channel.

We say that a channel $\mathsf{Ch}\colon D \to R$ with transition matrix $W$ is *symmetric* if the there exists a partition of the range as $R = R_1 \cup \cdots \cup R_n$ such that for all $i$ the sub-matrix $W[\cdot, R_i]$ induced by the rows in $R_i$ is strongly symmetric, i.e., all rows are permutations of each other, and all columns are permutations of each other.

Algorithms. We sometimes describe a transform (for example an encryption function or an adversary) algorithmically. In this view, a transform $T$ takes input $X$ and coins $R$ to deterministically return an output $Y \leftarrow T(X;R)$. By $Y \leftarrow\$\, T(X)$ we mean that we pick $R$ at random and let $Y \leftarrow T(X;R)$. The probability that a particular $Y$ is produced by this process is $W[X,Y]$ where $W$ is the transition probability matrix associated to transform $T$. As an example, we could specify $T$ by saying that on input $X$ and coins $R$, these being strings of the same length $m$, it returns $X \oplus R$. Then $Y \leftarrow\$\, T(X)$ means that we pick $R$ at random and let $Y = X \oplus R$, and the transition probability matrix $W$ is a $2^m$ by $2^m$ matrix all of whose entries equal $2^{-m}$. If $S$ is a (finite) set then $s \leftarrow\$\, S$ denotes the operation of picking a point at random from $S$ and denoting it $s$.

# 3 Security metrics

## 3.1 Encryption functions and schemes

An *encryption function* is a transform $\mathcal{E}\colon \{0,1\}^m \to \{0,1\}^c$ where $m$ is the message length and $c$ is the sender ciphertext length. The *rate* of $\mathcal{E}$ is $\mathbf{Rate}(\mathcal{E}) = m/c$. If $\mathsf{ChR}\colon \{0,1\}^c \to \{0,1\}^d$ is a receiver channel then a *decryption function* for $\mathcal{E}$ over $\mathsf{ChR}$ is a transform $\mathcal{D}\colon \{0,1\}^d \to \{0,1\}^m$ whose decryption error $\mathbf{DE}(\mathcal{E};\mathcal{D};\mathsf{ChR})$ is defined as the maximum, over all $M \in \{0,1\}^m$, of $\Pr[\mathcal{D}(\mathsf{ChR}(\mathcal{E}(M))) \neq M]$.

An *encryption scheme* $\overline{\mathcal{E}} = \{\mathcal{E}_k\}_{k\in\mathbb{N}}$ is a family of encryption functions where $\mathcal{E}_k\colon \{0,1\}^{m(k)} \to \{0,1\}^{c(k)}$ for functions $m,c\colon \mathbb{N} \to \mathbb{N}$ called the message length and sender ciphertext lengths of the scheme. The *rate* of $\overline{\mathcal{E}}$ is the function $\mathbf{Rate}_{\overline{\mathcal{E}}}\colon \mathbb{N} \to \mathbb{R}$ defined by $\mathbf{Rate}_{\overline{\mathcal{E}}}(k) = \mathbf{Rate}(\mathcal{E}_k)$ for all $k \in \mathbb{N}$. Suppose $\overline{\mathsf{ChR}} = \{\mathsf{ChR}_k\}_{k\in\mathbb{N}}$ is a family of receiver channels where $\mathsf{ChR}_k\colon \{0,1\}^{c(k)} \to \{0,1\}^{d(k)}$. Then a *decryption scheme* for $\overline{\mathcal{E}}$ over $\overline{\mathsf{ChR}}$ is a family $\overline{\mathcal{D}} = \{\mathcal{D}_k\}_{k\in\mathbb{N}}$ where $\mathcal{D}_k\colon \{0,1\}^{d(k)} \to \{0,1\}^{m(k)}$ is a decryption function for $\mathcal{E}_k$ over $\mathsf{ChR}_k$. We say that $\overline{\mathcal{D}}$ is a correct decryption scheme for $\overline{\mathcal{E}}$ relative to $\overline{\mathsf{ChR}}$ if $\lim_{k\to\infty} \mathbf{DE}(\mathcal{E}_k;\mathcal{D}_k;\mathsf{ChR}_k) = 0$. We say that encryption scheme $\overline{\mathcal{E}}$ is decryptable relative to $\overline{\mathsf{ChR}}$ if there exists a correct decryption scheme for $\overline{\mathcal{E}}$ relative to $\overline{\mathsf{ChR}}$.

This standard requirement from the I&C literature is, however, weak in that the rate at which the decryption error approaches zero could be very slow. For example, it would be met when $\mathbf{DE}(\mathcal{E}_k; \mathcal{D}_k; \mathsf{ChR}_k) = 1/\lg k$. We propose a stronger requirement, namely that the decryption error vanish exponentially in $k$. Thus we say that $\overline{\mathcal{D}}$ is a correct strong decryption scheme for $\overline{\mathcal{E}}$ relative to $\overline{\mathsf{ChR}}$ if there are constants $d, \epsilon > 0$ such that $-\lg(\mathbf{DE}(\mathcal{E}_k; \mathcal{D}_k; \mathsf{ChR}_k)) \geq dk^\epsilon$ for all $k \in \mathbb{N}$. We say that encryption scheme $\overline{\mathcal{E}}$ is strongly decryptable relative to $\overline{\mathsf{ChR}}$ if there exists a correct strong decryption scheme for $\overline{\mathcal{E}}$ relative to $\overline{\mathsf{ChR}}$.

We say that a family $\{\mathcal{S}_k\}_{k \in \mathbb{N}}$ (eg. an encryption or decryption scheme) is polynomial-time computable if there is a polynomial time computable function which on input $1^k$ (the unary representation of $k$) and $x$ returns $\mathcal{S}_k(x)$.

### 3.2 Security metrics

We are interested in measuring the security providing by an encryption function $\mathcal{E}\colon \{0,1\}^m \to \{0,1\}^c$ relative to an adversary channel $\mathsf{ChA}\colon \{0,1\}^c \to \{0,1\}^d$. (The receiver channel is not relevant to security.) A *security metric* xs associates to $\mathcal{E}; \mathsf{ChA}$ a real number $\mathbf{Adv}^{\mathrm{xs}}(\mathcal{E}; \mathsf{ChA})$. Intuitively, the latter is the amount of "information" about the message $\mathsf{M}$, as measured by metric xs, that is present in the adversary ciphertext $\mathsf{ChA}(\mathcal{E}(\mathsf{M}))$. The smaller this number, the more secure is $\mathcal{E}; \mathsf{ChA}$ according to the metric in question. The metrics we will define are semantic security (ss), distinguishing security (ds), mutual-information security (mis) and mutual-information security for random messages (mis-r).

In cryptography it is customary to measure security in bits, saying that $\mathcal{E}; \mathsf{ChA}$ has $s$ bits of xs-security if $\mathbf{Adv}^{\mathrm{xs}}(\mathcal{E}; \mathsf{ChA}) \leq 2^{-s}$. There is no formal definition of an encryption function being "secure" or "insecure," security rather being quantitative, given by a certain number of bits. A qualitative definition of "secure," meaning one under which something is secure or not secure, may only be made asymptotically, meaning for schemes. We say that encryption scheme $\overline{\mathcal{E}} = \{\mathcal{E}_k\}_{k \in \mathbb{N}}$ is XS-secure relative to $\overline{\mathsf{ChA}} = \{\mathsf{ChA}_k\}_{k \in \mathbb{N}}$ if $\lim_{k \to \infty} \mathbf{Adv}^{\mathrm{xs}}(\mathcal{E}_k; \mathsf{ChA}_k) = 0$. We say $\overline{\mathcal{E}} = \{\mathcal{E}_k\}_{k \in \mathbb{N}}$ is strongly XS-secure relative to $\overline{\mathsf{ChA}} = \{\mathsf{ChA}_k\}_{k \in \mathbb{N}}$ if there are constants $\epsilon, d > 0$ such that $-\lg \mathbf{Adv}^{\mathrm{xs}}(\mathcal{E}_k; \mathsf{ChA}_k) \geq dk^\epsilon$ for all $k \in \mathbb{N}$.

With encryption function $\mathcal{E}\colon \{0,1\}^m \to \{0,1\}^c$ and adversary channel $\mathsf{ChA}\colon \{0,1\}^c \to \{0,1\}^d$ fixed, we now move to defining the above-mentioned metrics.

### 3.3 Mutual-information metrics

We dub the metric in current use mutual-information security for random messages (mis-r). The corresponding advantage is defined via

$$\mathbf{Adv}^{\text{mis-r}}(\mathcal{E}; \mathsf{ChA}) \;=\; \mathbf{I}(\mathsf{U}; \mathsf{ChA}(\mathcal{E}(\mathsf{U}))) \tag{1}$$

where the random variable $\mathsf{U}$ is uniformly distributed over $\{0,1\}^m$. The weakness of this metric is that it only considers uniformly distributed messages. Yet "real" messages are not uniformly distributed. Indeed, they may be drawn from small and structured spaces. They may be English text. They may take only values in some small and known set $S$. For example, they may be votes, where $S = \{0,1\}$, or scores on an exam, where $S = \{0, 1, \ldots, 100\}$. Mis-r security will thus not ensure security when encrypting the type of data that actually arises in applications. This leads us to define mutual-information security (mis) via

$$\mathbf{Adv}^{\mathrm{mis}}(\mathcal{E}; \mathsf{ChA}) \;=\; \max_{\mathsf{M}} \mathbf{I}(\mathsf{M}; \mathsf{ChA}(\mathcal{E}(\mathsf{M}))) \tag{2}$$

where the maximum is over all random variables $\mathsf{M}$ over $\{0,1\}^m$. This definition is what we call *message-distribution independent* in that security is required regardless of how messages are distributed. In this way, distributions that arise in applications are captured.

We remark that the importance of message-distribution independence of a metric was understood by Shannon [42]. His definition of perfect privacy did not assume uniformly distributed messages or, in fact, any particular distribution on messages, but held across all distributions. It is curious that this feature was dropped in defining a security metric for the wiretap channel. We have resurrected it.

In cryptography, the importance of message-distribution independence has been understood since [19] and is now ubiquitously viewed as necessary for a "good" definition. We will continue to require message-distribution independence in our subsequent metrics.

In the wiretap literature it is sometimes argued that the assumption of uniformly distributed messages is tenable because messages are compressed before transmission. But compression does not result in uniformly distributed messages. Compression is a deterministic, injective function and does not change the entropy.

At this point we have in mis a legitimate metric, but one whose underlying intuition is, at least for cryptographers, obscure. Mis-security measures the difference in the number of bits required to encode the message before and after seeing the ciphertext. Cryptographers have very different approaches and intuition with regard to security and we now turn to definitions based on those.

### 3.4 Semantic security

We define the ss advantage via

$$\mathbf{Adv}^{\mathrm{ss}}(\mathcal{E};\mathsf{ChA}) \;=\; \max_{f,\mathsf{M}} \left( \max_{\mathcal{A}} \Pr[\mathcal{A}(\mathsf{ChA}(\mathcal{E}(\mathsf{M}))) = f(\mathsf{M})] - \max_{\mathcal{S}} \Pr[\mathcal{S}(m) = f(\mathsf{M})] \right) . \tag{3}$$

The maximum is over all random variables $\mathsf{M}$ over $\{0,1\}^m$ and all transforms $f$ with domain $\{0,1\}^m$.

Think of the adversary $\mathcal{A}$, given the adversary ciphertext $\mathsf{ChA}(\mathcal{E}(\mathsf{M}))$, as trying to output some function $f$ of the message $\mathsf{M}$. In the simplest case $f$ is the identity function, so that the adversary is trying to recover the message. However, figuring out partial information about the message (rather than the entire message) should still be considered a violation of security. For example, the first bit of the message could be a vote that we want to hide, which would be captured by letting $f$ be the function that returns the first bit of its input. To ensure that no partial information leaks, we allow $f$ to be any function.

The term $\Pr[\mathcal{A}(\mathsf{ChA}(\mathcal{E}(\mathsf{M}))) = f(\mathsf{M})]$, maximized over all adversaries, represents the maximum possible probability that an adversary could output $f(\mathsf{M})$ given $\mathsf{ChA}(\mathcal{E}(\mathsf{M}))$. This by itself, however, is not a measure of its success, because knowledge of $f, \mathsf{M}$ entails that the adversary has some a priori probability of being able to output $f(\mathsf{M})$ even if did not see the ciphertext. The subtracted term accounts for this. It is the maximum, over all simulators $\mathcal{S}$, of the probability $\Pr[\mathcal{S}(m) = f(\mathsf{M})]$ that the simulator can figure out $f(\mathsf{M})$ given nothing but the length of the message and the implicit knowledge of $f, \mathsf{M}$ represented by the order of quantification. A simulator is (recall) simply a transform, the name chosen to allude to its role. The need to so adjust for a priori knowledge was recognized by Shannon [42] in the context of perfect privacy and Goldwasser and Micali in their definition of semantic security for public-key cryptosystems [19].

Finally, the outer maximum over $f$ means that we require no partial information to leak, regardless of message distribution.

This adversary-based formulation reflects the cryptographic approach and, with it, semantic security quite directly captures a strong and natural intuition, namely that you can't compute any function of the encrypted message with probability better than what you could guess if you never had the ciphertext. Having written it this way, however, we now remark that it can be re-written using information-theoretic quantities. Namely, we claim that

$$\mathbf{Adv}^{\mathrm{ss}}(\mathcal{E};\mathsf{ChA}) \;=\; \sup_{f,\mathsf{M}} \left( \mathbf{GP}(f(\mathsf{M}) | \mathsf{ChA}(\mathcal{E}(\mathsf{M}))) - \mathbf{GP}(f(\mathsf{M})) \right) \tag{4}$$

where the maximum is over all random variables $\mathsf{M}$ over $\{0,1\}^m$ and all transforms $f$ with domain $\{0,1\}^m$. The proof that Equations (3) and (4) define the same quantities is left as an exercise.

### 3.5 Distinguishing security

We define the ds advantage via

$$\mathbf{Adv}^{\rm ds}(\mathcal{E};\mathsf{ChA}) = \max_{\mathcal{A},M_0,M_1} 2\Pr[\mathcal{A}(M_0,M_1,\mathsf{ChA}(\mathcal{E}(M_{\mathsf{b}}))) = \mathsf{b}] - 1 \quad (5)$$
$$= \max_{M_0,M_1} \mathbf{SD}(\mathsf{ChA}(\mathcal{E}(M_0));\mathsf{ChA}(\mathcal{E}(M_1))) \; . \quad (6)$$

In Eq. (5), $\mathsf{b}$ is a random variable uniformly distributed over $\{0,1\}$. The maximization is over all $M_0, M_1 \in \{0,1\}^m$ (we stress these are strings, not random variables) and all adversaries $\mathcal{A}$. $\Pr[\mathcal{A}(M_0,M_1,\mathsf{ChA}(\mathcal{E}(M_{\mathsf{b}}))) = \mathsf{b}]$ is the probability that adversary $\mathcal{A}$, given $m$-bit messages $M_0, M_1$ and an adversary ciphertext emanating from $M_{\mathsf{b}}$, correctly identifies the random challenge bit $\mathsf{b}$. The a priori success probability being 1/2, the advantage is appropriately scaled. This advantage is equal to the statistical distance between the random variables $\mathsf{ChA}(\mathcal{E}(M_0))$ and $\mathsf{ChA}(\mathcal{E}(M_1))$ as per Eq. (6).

In this definition, one only considers the subclass of message distributions whose support is a set of at most two equi-probable messages. Yet it will be equivalent to semantic security. The value of ds is that it is easier to work with than ss yet equivalent to it.

# 4 Relations

We establish the relations summarized in Figure 2.

### 4.1 DS is equivalent to SS

The following says that SS and DS are equivalent up to a small constant factor. The proof is an extension of the classical ones in computational cryptography.

**Theorem 4.1 [DS $\leftrightarrow$ SS]** *Let $\mathcal{E}$: $\{0,1\}^m \to \{0,1\}^c$ be an encryption algorithm and* $\mathsf{ChA}$ *an adversary channel. Then* $\mathbf{Adv}^{\rm ss}(\mathcal{E};\mathsf{ChA}) \le \mathbf{Adv}^{\rm ds}(\mathcal{E};\mathsf{ChA}) \le 2\cdot\mathbf{Adv}^{\rm ss}(\mathcal{E};\mathsf{ChA})$. ▌

**Proof:** Let $\mathcal{A}_{\rm ss}$ be an adversary attacking the SS security of $\mathcal{E};\mathsf{ChA}$. We construct $\mathcal{A}_{\rm ds}$ attacking the DS security of $\mathcal{E};\mathsf{ChA}$ as follows. On inputs $M_0, M_1$ and adversary ciphertext $C$, adversary $\mathcal{A}_{\rm ds}$ runs $\mathcal{A}_{\rm ss}$ on input $C$ to get a value $v$. If $v = f(M_1)$ it outputs 1, else it outputs 0. Let $\mathsf{M}_0, \mathsf{M}_1$ be distributed identically to $\mathsf{M}$ but independently of each other. Then

$$\Pr[\mathcal{A}_{\rm ds}(\mathsf{M}_0,\mathsf{M}_1,\mathsf{ChA}(\mathcal{E}(\mathsf{M}_1))) = 1] = \Pr[\mathcal{A}_{\rm ss}(\mathsf{ChA}(\mathcal{E}(\mathsf{M}))) = f(\mathsf{M})]$$
$$\Pr[\mathcal{A}_{\rm ds}(\mathsf{M}_0,\mathsf{M}_1,\mathsf{ChA}(\mathcal{E}(\mathsf{M}_0))) = 1] \le \max_{\mathcal{S}} \Pr[\mathcal{S}(m) = f(\mathsf{M})] \; .$$

Subtracting, we get

$$\begin{aligned}
&\Pr[\mathcal{A}_{\rm ss}(\mathsf{ChA}(\mathcal{E}(\mathsf{M}))) = f(\mathsf{M})] - \max_{\mathcal{S}} \Pr[\mathcal{S}(m) = f(\mathsf{M})] \\
&\le \Pr[\mathcal{A}_{\rm ds}(\mathsf{M}_0,\mathsf{M}_1,\mathsf{ChA}(\mathcal{E}(\mathsf{M}_1))) = 1] - \Pr[\mathcal{A}_{\rm ds}(\mathsf{M}_0,\mathsf{M}_1,\mathsf{ChA}(\mathcal{E}(\mathsf{M}_0))) = 1] \\
&= 2\Pr[\mathcal{A}_{\rm ds}(\mathsf{M}_0,\mathsf{M}_1,\mathsf{ChA}(\mathcal{E}(\mathsf{M}_{\mathsf{b}}))) = \mathsf{b}] - 1 \\
&\le \max_{M_0,M_1} 2\Pr[\mathcal{A}_{\rm ds}(M_0,M_1,\mathsf{ChA}(\mathcal{E}(M_{\mathsf{b}}))) = \mathsf{b}] - 1 \; .
\end{aligned}$$

Taking the max over all adversaries and all $\mathsf{M}$ yields $\mathbf{Adv}^{\rm ss}(\mathcal{E};\mathsf{ChA}) \le \mathbf{Adv}^{\rm ds}(\mathcal{E};\mathsf{ChA})$.

We now prove $\mathbf{Adv}^{\rm ds}(\mathcal{E};\mathsf{ChA}) \le 2\cdot\mathbf{Adv}^{\rm ss}(\mathcal{E};\mathsf{ChA})$. For any (wlog) distinct $M_0, M_1 \in \{0,1\}^m$ let $\mathsf{M}^{M_0,M_1}$ denote the random variable that is uniformly distributed over $\{M_0, M_1\}$. Then

$$\begin{aligned}
\frac{1}{2}\mathbf{Adv}^{\rm ds}(\mathcal{E};\mathsf{ChA}) &= \max_{M_0,M_1}\max_{\mathcal{A}_{\rm ds}}\left(\Pr[\mathcal{A}_{\rm ds}(M_0,M_1,\mathsf{ChA}(\mathcal{E}(M_{\mathsf{b}}))) = \mathsf{b}] - \frac{1}{2}\right) \\
&= \max_{M_0,M_1}\max_{\mathcal{A}_{\rm ss}}\left(\Pr[\mathcal{A}_{\rm ss}(\mathsf{ChA}(\mathcal{E}(\mathsf{M}^{M_0,M_1}))) = \mathsf{M}^{M_0,M_1}] - \max_{\mathcal{S}}\Pr[\mathcal{S}(m) = \mathsf{M}^{M_0,M_1}]\right) \\
&\le \mathbf{Adv}^{\rm ss}(\mathcal{E};\mathsf{ChA}) \; ,
\end{aligned}$$

where the max is over all distinct $M_0, M_1 \in \{0,1\}^m$. This simply reflects the fact that on a distribution over two distinct messages $M_0, M_1$, if $\mathsf{b}$ is random, finding $\mathsf{b}$ and finding $M_{\mathsf{b}}$ are equivalent tasks. ∎

A corollary of this is that an encryption scheme $\overline{\mathcal{E}} = \{\mathcal{E}_k\}_{k\in\mathbb{N}}$ is (strongly) SS-secure relative to a family of channels $\overline{\mathsf{ChA}} = \{\mathsf{ChA}_k\}_{k\in\mathbb{N}}$ if and only if it is (respectively, strongly) DS-secure relative to the same family of channels. We stress that Theorem 4.1, and this corollary, hold for all channels, meaning channels specified by arbitrary transforms. This certainly includes binary channels such as the one induced by the binary symmetric channel or other symmetric channels. However, it also includes many other channels. An (implicit) assumption, however is that successive uses of the channel are independent, meaning if multiple messages are encrypted then the different uses of $\mathsf{ChA}$ are independent.

The equivalence between DS and SS is helpful because DS is more analytically tractable than SS, and we exploit it when we come to the design of solutions.

### 4.2 MIS versus DS

Neither direction of the equivalence between MIS and DS is trivial. Going back to Eq. (2), $\mathbf{H}(\mathsf{M})$ is the minimum number of bits to encode a message from $\mathsf{M}$ and $\mathbf{H}(\mathsf{M} \mid \mathsf{ChA}(\mathcal{E}(\mathsf{M})))$ is the minimum number of bits to encode the message after seeing the ciphertext, so mi security measures the reduction in encoding length of messages allowed by seeing the ciphertext. It is not intuitively clear how this encoding length intuition relates to statistical distance or semantic security. From the analytic perspective, letting $\mathsf{C} = \mathsf{ChA}(\mathcal{E}(\mathsf{M}))$, the mutual information $\mathbf{I}(\mathsf{M};\mathsf{C})$ involved in Eq. (2) is

$$-\sum_M \Pr[\mathsf{M}=M]\lg\Pr[\mathsf{M}=M] + \sum_C \Pr[\mathsf{C}=C]\sum_M \Pr[\mathsf{M}=M|\mathsf{C}=C]\cdot\lg\Pr[\mathsf{M}=M|\mathsf{C}=C]\ .$$

This is a complex expression that looks quite different from statistical distance, in particular because of the logarithms. Additionally we need to consider the maximum over $\mathsf{M}$. We will first show that MIS implies DS and then by an entirely different technique that DS implies MIS.

### 4.3 MIS implies DS

We begin by recalling that the KL divergence is a distance metric for probability distributions $P, Q$ defined by $\mathbf{D}(P;Q) = \sum_x P(x)\lg P(x)/Q(x)$. Let $\mathsf{M}, \mathsf{C}$ be random variables. Probability distributions $J_{M,C}, I_{M,C}$ are defined for all $M, C$ by $J_{\mathsf{M},\mathsf{C}}(M,C) = \Pr[\,\mathsf{M}=M, \mathsf{C}=C\,]$ and $I_{\mathsf{M},\mathsf{C}}(M,C) = \Pr[\,\mathsf{M}=M\,]\cdot\Pr[\,\mathsf{C}=C\,]$. Thus $J_{\mathsf{M},\mathsf{C}}$ is the joint distribution of $\mathsf{M}$ and $\mathsf{C}$, while $I_{\mathsf{M},\mathsf{C}}$ is the "independent" or product distribution. The following lemma recasts mutual information in terms of the KL divergence between the joint and independent distributions.

**Lemma 4.2** *Let* $\mathsf{M}, \mathsf{C}$ *be random variables. Then* $\mathbf{I}(\mathsf{M};\mathsf{C}) = \mathbf{D}(J_{\mathsf{M},\mathsf{C}}; I_{\mathsf{M},\mathsf{C}})$. ∎

The proof is standard and recalled for completeness in Appendix B.1. We have taken this path in the hope of exploiting Pinsker's inequality —from [38] with the tight constant from [12]— which lower bounds the KL divergence between two distributions in terms of their statistical distance:

**Lemma 4.3 [Pinsker's Inequality]** *Let* $P, Q$ *be probability distributions. Then* $\mathbf{D}(P;Q) \geq 2\cdot\mathbf{SD}(P;Q)^2$. ∎

At this point, from Lemmas 4.2 and 4.3, letting $\mathsf{C} = \mathsf{ChA}(\mathcal{E}(\mathsf{M}))$ denote the adversary ciphertext, we have

$$\mathbf{Adv}^{\mathrm{mis}}(\mathcal{E};\mathsf{ChA}) \;=\; \max_{\mathsf{M}} \mathbf{D}(J_{\mathsf{M},\mathsf{C}} I_{\mathsf{M},\mathsf{C}}) \;\geq\; 2\cdot\max_{\mathsf{M}} \mathbf{SD}(J_{\mathsf{M},\mathsf{C}}; I_{\mathsf{M},\mathsf{C}})^2\ . \tag{7}$$

We would now like to connect the RHS of Eq. (7) to $\mathbf{Adv}^{\mathrm{ds}}(\mathcal{E};\mathsf{ChA})$ so as to lower bound the mis advantage in terms of the ds advantage. (The upper bound will need a completely different approach, Pinsker's inequality being of no use for it.) It turns out that one can show that

$$\max_{\mathsf{M}} \mathbf{SD}(J_{\mathsf{M},\mathsf{C}}; I_{\mathsf{M},\mathsf{C}}) \;\leq\; \mathbf{Adv}^{\mathrm{ds}}(\mathcal{E};\mathsf{ChA})\ . \tag{8}$$

This, inequality, unfortunately, goes the wrong way, in the sense that, combining it with Eq. (7) does not lower bound mis in terms of ds. The observation that gets around this is that the inequality of Eq. (8) becomes an equality when one restricts attention to $\mathsf{M}$ distributed uniformly over a set of two messages. More precisely:

**Lemma 4.4** *Let $M_0, M_1 \in \{0,1\}^m$ and let $\mathsf{M}$ be uniformly distributed over $\{M_0, M_1\}$. Let $g\colon \{0,1\}^m \to \{0,1\}^c$ be a transform and let $\mathsf{C} = g(\mathsf{M})$. Then*

$$\mathbf{SD}(J_{\mathsf{M},\mathsf{C}}; I_{\mathsf{M},\mathsf{C}}) \;=\; \frac{1}{2} \cdot \mathbf{SD}(g(M_0); g(M_1)) \;. \;\blacksquare$$

**Proof:** We have

$$\begin{aligned}
\mathbf{SD}(J_{\mathsf{M},\mathsf{C}}; I_{\mathsf{M},\mathsf{C}}) &= \frac{1}{2}\sum_{m,c} |J_{\mathsf{M},\mathsf{C}}(m,c) - I_{\mathsf{M},\mathsf{C}}(m,c)| \\
&= \frac{1}{2}\sum_{m,c} \left|P_{\mathsf{M}}(m)\cdot P_{\mathsf{C}|\mathsf{M}=m}(c) - P_{\mathsf{M}}(m)\cdot P_{\mathsf{C}}(c)\right| \\
&= \frac{1}{2}\sum_{m} P_{\mathsf{M}}(m) \sum_{c} \left|P_{\mathsf{C}|\mathsf{M}=m}(c) - P_{\mathsf{C}}(c)\right| \\
&= \frac{1}{2}\sum_{m} P_{\mathsf{M}}(m) \sum_{c} \left|P_{\mathsf{C}|\mathsf{M}=m}(c) - \sum_{m'} P_{\mathsf{C}|\mathsf{M}=m'}(c)\cdot P_{\mathsf{M}}(m')\right| \\
&= \frac{1}{2}\sum_{m} P_{\mathsf{M}}(m) \sum_{c} \left|\sum_{m'} (P_{\mathsf{C}|\mathsf{M}=m}(c) - P_{\mathsf{C}|\mathsf{M}=\mathsf{m}'}(c))\cdot P_{\mathsf{M}}(m')\right| \;.
\end{aligned}$$

So far we have not used the assumption that $\mathsf{M}$ takes only two values, each with probability 1/2. Let $\alpha(m_0) = m_1$ and $\alpha(m_1) = m_0$. Then the sum over $m'$ above has only one candidate non-zero term, namely the one corresponding to $m' = \alpha(m)$. So the above equals

$$\begin{aligned}
&\frac{1}{2}\sum_{m} P_{\mathsf{M}}(m) \sum_{c} \left|(P_{\mathsf{C}|\mathsf{M}=m}(c) - P_{\mathsf{C}|\mathsf{M}=\alpha(m)}(c))\cdot P_{\mathsf{M}}(\alpha(m))\right| \\
&= \frac{1}{2}\sum_{m} \frac{1}{2} \sum_{c} \frac{1}{2} \left|P_{\mathsf{C}|\mathsf{M}=M_0}(c) - P_{\mathsf{C}|\mathsf{M}=M_1}(c)\right| \\
&= \frac{1}{4}\cdot 2 \cdot \frac{1}{2} \cdot \sum_{c} \left|P_{\mathsf{C}|\mathsf{M}=M_0}(c) - P_{\mathsf{C}|\mathsf{M}=M_1}(c)\right| \\
&= \frac{1}{2}\cdot \mathbf{SD}(P_{\mathsf{C}|\mathsf{M}=M_0}; P_{\mathsf{C}|\mathsf{M}=M_1}) \\
&= \frac{1}{2}\cdot \mathbf{SD}(g(M_0); g(M_1)) \;.
\end{aligned}$$

This completes the proof. $\blacksquare$

Now we combine the lemmas to prove that MIS implies DS:

**Theorem 4.5 [MIS $\to$ DS]** *Let $\mathcal{E}\colon \{0,1\}^m \to \{0,1\}^c$ be an encryption algorithm and $\mathsf{ChA}$ an adversary channel. Then $\mathbf{Adv}^{\mathrm{ds}}(\mathcal{E}; \mathsf{ChA}) \le \sqrt{2\cdot \mathbf{Adv}^{\mathrm{mis}}(\mathcal{E}; \mathsf{ChA})}$.* $\blacksquare$

**Proof:** We have already applied Lemmas 4.2 and 4.3 to get Eq. (7). Now let $M_0, M_1 \in \{0,1\}^m$ be messages for which $\mathbf{SD}(\mathsf{ChA}(\mathcal{E}(M_0)); \mathsf{ChA}(\mathcal{E}(M_1)))$ equals $\mathbf{Adv}^{\mathrm{ds}}(\mathcal{E}; \mathsf{ChA})$. Let $\mathsf{M}$ be uniformly distributed over $\{M_0, M_1\}$. Since $\mathsf{M}$ is in the scope of the max in (7), we can apply Lemma 4.4 to see that the RHS of Eq. (7) is at least

$$2\cdot \mathbf{SD}(J_{\mathsf{M},\mathsf{C}}; I_{\mathsf{M},\mathsf{C}})^2 \;=\; 2\cdot\frac{1}{4}\cdot \mathbf{SD}(\mathsf{ChA}(\mathcal{E}(M_0)); \mathsf{ChA}(\mathcal{E}(M_1))^2 \;=\; \frac{1}{2}\cdot \mathbf{Adv}^{\mathrm{ds}}(\mathcal{E}; \mathsf{ChA})^2 \;,$$

where again $\mathsf{C} = \mathsf{ChA}(\mathcal{E}(\mathsf{M}))$. $\blacksquare$

A corollary of Theorem 4.5 is that if an encryption scheme $\overline{\mathcal{E}} = \{\mathcal{E}_k\}_{k\in\mathbb{N}}$ is (strongly) MIS-secure relative to a family of channels $\overline{\mathsf{ChA}} = \{\mathsf{ChA}_k\}_{k\in\mathbb{N}}$ then it is also (respectively, strongly) DS-secure relative to the same family of channels. Again, Theorem 4.5, and this corollary, hold for all channels, not just binary ones. On the quantitative side, however, Theorem 4.5 says that $s$ bits of mis security imply $(s-1)/2 \approx s/2$ bits of ds security. We do not know whether this bound is tight.

### 4.4 DS implies MIS

The proof is underlain by the following general lemma that bounds the difference in entropy between two distributions in terms of their statistical distance:

**Lemma 4.6** *Let $P, Q$ be probability distributions. Let $N = |\textsc{supp}(P) \cup \textsc{supp}(Q)|$ and $\epsilon = \mathbf{SD}(P;Q)$. Then $\mathbf{H}(P) - \mathbf{H}(Q) \le 2\epsilon \cdot \lg(N/\epsilon)$.* ▮

To prove this we need the following, which appears as Equation (16.24) in [11, Section 16.3]:

**Lemma 4.7** *Let $p, x \in [0, 1/2]$ and assume $p + x \le 1/2$. Then $|h(p+x) - h(p)| \le h(x)$.* ▮

The following is similar to the proof of [11, Theorem 16.3.2].

**Proof of Lemma 4.6:** Let $\delta(y) = |P(y)/2 - Q(y)/2| \le 1/2$ for all $y$. Then

$$\begin{aligned}
\mathbf{H}(P) - \mathbf{H}(Q) &= \sum_y h(P(y)) - h(Q(y)) \;=\; \sum_y P(y)\lg\frac{1}{P(y)} - Q(y)\lg\frac{1}{Q(y)} \\
&= \sum_y P(y)\lg\frac{1}{P(y)/2} - Q(y)\lg\frac{1}{Q(y)/2} \;=\; 2\cdot\sum_y \frac{P(y)}{2}\lg\frac{1}{P(y)/2} - \frac{Q(y)}{2}\lg\frac{1}{Q(y)/2} \\
&= 2\cdot\sum_y h(P(y)/2) - h(Q(y)/2) \;\le\; 2\cdot\sum_y |h(P(y)/2) - h(Q(y)/2)| \;.
\end{aligned}$$

If $P(y) \ge Q(y)$ then $P(y)/2 + \delta(y) = Q(y)/2 \le 1/2$ so we can apply Lemma 4.7 to get

$$|h(P(y)/2) - h(Q(y)/2)| \;=\; |h(P(y)/2) - h(P(y)/2 + \delta(y))| \;\le\; h(\delta(y)) \;.$$

On the other hand if $Q(y) \ge P(y)$ then $Q(y)/2 + \delta(y) = P(y)/2 \le 1/2$ so we can apply Lemma 4.7 to get

$$|h(P(y)/2) - h(Q(y)/2)| \;=\; |h(Q(y)/2 + \delta(y)) - h(Q(y)/2)| \;\le\; h(\delta(y)) \;.$$

Continuing the above we have

$$\begin{aligned}
\mathbf{H}(P) - \mathbf{H}(Q) &\le 2\cdot\sum_y h(\delta(y)) \;=\; 2\cdot\sum_y -\delta(y)\lg\delta(y) \\
&= 2\epsilon\cdot\sum_y -\frac{\delta(y)}{\epsilon}\lg\delta(y) \;=\; 2\epsilon\cdot\sum_y -\frac{\delta(y)}{\epsilon}\lg\frac{\delta(y)}{\epsilon} - \frac{\delta(y)}{\epsilon}\lg\epsilon \\
&= 2\epsilon\cdot\sum_y h(\delta(y)/\epsilon) - 2\epsilon\cdot\sum_y \frac{\delta(y)}{\epsilon}\lg\epsilon \;=\; 2\epsilon\cdot\sum_y h(\delta(y)/\epsilon) - 2\lg\epsilon\cdot\sum_y \delta(y) \\
&= 2\epsilon\cdot\sum_y h(\delta(y)/\epsilon) - 2\epsilon\lg\epsilon \;\le\; 2\epsilon\cdot\lg N - 2\epsilon\lg\epsilon \\
&= 2\epsilon\cdot\lg\frac{N}{\epsilon}
\end{aligned}$$

as claimed. ▮

To exploit this, we define the *pairwise statistical distance* between random variables $\mathsf{M}, \mathsf{C}$ via

$$\mathbf{PSD}(\mathsf{M};\mathsf{C}) \;=\; \max_{M_0, M_1} \mathbf{SD}(P_{\mathsf{C}|\mathsf{M}=M_0}; P_{\mathsf{C}|\mathsf{M}=M_1}) \;,$$

where the maximum is over all $M_0, M_1 \in \textsc{supp}(P_{\mathsf{M}})$. We now have:

**Lemma 4.8** *Let* $\mathsf{M}, \mathsf{C}$ *be random variables. Then* $\mathbf{SD}(P_{\mathsf{C}}; P_{\mathsf{C}|\mathsf{M}=M}) \leq \mathbf{PSD}(\mathsf{M}; \mathsf{C})$ *for any* $M$*.*

**Proof:** We have

$$\begin{aligned}
\frac{1}{2}\sum_{c} \left|P_{\mathsf{C}|\mathsf{M}=m}(c) - P_{\mathsf{C}}(c)\right| &= \frac{1}{2}\sum_{c} \left|P_{\mathsf{C}|\mathsf{M}=m}(c) - \sum_{m'} P_{\mathsf{C}|\mathsf{M}=m'}(c) \cdot P_{\mathsf{M}}(m')\right| \\
&= \frac{1}{2}\sum_{c} \left|\sum_{m'} (P_{\mathsf{C}|\mathsf{M}=m}(c) - P_{\mathsf{C}|\mathsf{M}=m'}(c)) \cdot P_{\mathsf{M}}(m')\right| \\
&\leq \frac{1}{2}\sum_{c}\sum_{m'} \left|P_{\mathsf{C}|\mathsf{M}=m}(c) - P_{\mathsf{C}|\mathsf{M}=m'}(c)\right| \cdot P_{\mathsf{M}}(m') \\
&= \frac{1}{2}\sum_{m'} P_{\mathsf{M}}(m') \sum_{c} \left|P_{\mathsf{C}|\mathsf{M}=m}(c) - P_{\mathsf{C}|\mathsf{M}=m'}(c)\right| \\
&\leq \sum_{m'} P_{\mathsf{M}}(m') \cdot \mathbf{PSD}(\mathsf{M}; \mathsf{C}) \\
&= \mathbf{PSD}(\mathsf{M}; \mathsf{C})
\end{aligned}$$

as claimed. ▮

We now show that DS implies MIS.

**Theorem 4.9 [DS → MIS]** *Let* $\mathcal{E}\colon \{0,1\}^m \to \{0,1\}^c$ *be an encryption algorithm and* $\mathsf{ChA}$ *an adversary channel. Let* $\epsilon = \mathbf{Adv}^{\mathrm{ds}}(\mathcal{E}; \mathsf{ChA})$*. Then* $\mathbf{Adv}^{\mathrm{mis}}(\mathcal{E}; \mathsf{ChA}) \leq 2\epsilon \cdot \lg(2^c/\epsilon)$*.* ▮

**Proof:** Let $f(x) = \min(2x\lg(2^c/x), 1)$. Let $\mathsf{C} = \mathsf{ChA}(\mathcal{E}(\mathsf{M}))$ be the adversary ciphertext. Then

$$\mathbf{I}(\mathsf{M}; \mathsf{C}) = \mathbf{I}(\mathsf{C}; \mathsf{M}) = \mathbf{H}(P_{\mathsf{C}}) - \textstyle\sum_M P_{\mathsf{M}}(M) \cdot \mathbf{H}(P_{\mathsf{C}|\mathsf{M}=M}) = \textstyle\sum_M P_{\mathsf{M}}(M) \cdot \big(\mathbf{H}(P_{\mathsf{C}}) - \mathbf{H}(P_{\mathsf{C}|\mathsf{M}=M})\big) .$$

Fix a message $M$ such that $\mathbf{H}(P_{\mathsf{C}}) - \mathbf{H}(P_{\mathsf{C}|\mathsf{M}=M})$ equals the maximum, over all $M'$, of $\mathbf{H}(P_{\mathsf{C}}) - \mathbf{H}(P_{\mathsf{C}|\mathsf{M}=M'})$ and let $x = \mathbf{SD}(P_{\mathsf{C}}; P_{\mathsf{C}|\mathsf{M}=M})$. From Lemma 4.6 and the above we get

$$\mathbf{I}(\mathsf{M}; \mathsf{C}) \leq \mathbf{H}(P_{\mathsf{C}}) - \mathbf{H}(P_{\mathsf{C}|\mathsf{M}=M}) \leq f(x) .$$

Let $y = \mathbf{PSD}(\mathsf{M}; \mathsf{C})$. Lemma 4.8 says $x \leq y$. The function $f$ has the property that $x \leq y$ implies $f(x) \leq f(y)$. So the above is at most $f(y)$. Finally, take the maximum over all $\mathsf{M}$ on both sides. ▮

We would like as a corollary of Theorem 4.9 to say that if an encryption scheme $\overline{\mathcal{E}} = \{\mathcal{E}_k\}_{k\in\mathbb{N}}$ is (strongly) DS-secure relative to a family of channels $\overline{\mathsf{ChA}} = \{\mathsf{ChA}_k\}_{k\in\mathbb{N}}$ then it is also (respectively, strongly) MIS-secure relative to the same family of channels. Theorem 4.9 does not imply this for all encryption schemes but it does usually. Let $m, c$ be the message length and sender ciphertext length of $\overline{\mathcal{E}}$. Let $\epsilon(k) = \mathbf{Adv}^{\mathrm{ds}}(\mathcal{E}_k; \mathsf{ChA}_k)$ and $\delta(k) = 2\epsilon(k)\lg(2^{c(k)}/\epsilon(k))$. Usually $m, c$ are bounded by polynomials in $k$ while $\epsilon(k)$ decreases exponentially in $k$. In this case, we have the desired conclusions. We do under other conditions as well. The case where we would not is the unnatural one that $c(k)$ is extremely fast growing, for example exponential in $k$, yet $\epsilon(k)$ decreases no faster than exponentially.

We do not know whether the bound of Theorem 4.9 is tight. The following, however, says that the bound of Lemma 4.6 is tight up to a constant factor. This means that improving the bound of Theorem 4.9 would require a different approach.

**Proposition 4.10** *Let* $n > k \geq 1$ *be integers. Let* $\epsilon = 2^{-k}$ *and* $N = 1 + \epsilon 2^n$*. Then there are distributions* $P, Q$ *with* $|\textsc{supp}(P) \cup \textsc{supp}(Q)| = N$ *and* $\mathbf{SD}(P; Q) = \epsilon$ *and* $\mathbf{H}(P) - \mathbf{H}(Q) \geq 0.5 \cdot \epsilon \cdot \lg(N/\epsilon)$*.* ▮

**Proof:** Let $S = \{0, 1, \ldots, N-1\}$. Let $P, Q\colon S \to [0,1]$ be defined as follows. Let $P(i) = 2^{-n}$ for $i \geq 1$ and $P(0) = 1 - \epsilon$. Let $Q(i) = 0$ for $i \geq 1$ and $Q(0) = 1$. Then

$$\mathbf{SD}(P; Q) \;=\; \frac{1}{2} \cdot (\epsilon + (N-1) \cdot 2^{-n}) \;=\; \frac{1}{2} \cdot (\epsilon + \epsilon 2^n \cdot 2^{-n}) \;=\; \epsilon .$$

On the other hand

$$\mathbf{H}(P) - \mathbf{H}(Q) \;=\; \big(\epsilon 2^n \cdot h(2^{-n}) + h(1-\epsilon)\big) - h(1) \;=\; \epsilon n + h(1-\epsilon) \;\geq\; \epsilon n\ .$$

However

$$\frac{\epsilon}{2} \cdot \lg \frac{N}{\epsilon} \;=\; \frac{\epsilon}{2} \cdot \lg \frac{1+\epsilon 2^n}{\epsilon} \;\leq\; \frac{\epsilon}{2} \cdot \lg \frac{2\epsilon 2^n}{\epsilon} \;=\; \frac{\epsilon}{2} \cdot (n+1) \;\leq\; \epsilon n$$

which proves the claim. ▮

### 4.5 MIS-R does not imply DS

At this point we have justified all the numbered implication arrows in Figure 2. The un-numbered implication **MIS** $\rightarrow$ **MIS-R** is trivial. Let us turn to the separation **MIS-R** $\not\rightarrow$ **MIS**. To justify it we need to exhibit an encryption scheme $\overline{\mathcal{E}}$ and a family $\overline{\mathsf{ChA}}$ of adversary channels such that $\overline{\mathcal{E}}$ is (strongly) MIS-R-secure relative to $\overline{\mathsf{ChA}}$ but not (respectively, strongly) DS-secure relative to $\overline{\mathsf{ChA}}$.

This is easy to do with a contrived choice of $\overline{\mathsf{ChA}}$. (Have the channel faithfully transmit inputs $0^m$ and $1^m$ and be very noisy on other inputs. Then **MIS** fails because the adversary has high advantage when the message takes on only values $0^m, 1^m$ but MIS-R-security holds since these messages are unlikely.) This however is not very convincing because of the obscure nature of the channel. We will instead give an example where the channel is the binary symmetric one. The counter-example is constructed by starting with a scheme that is MIS-R-secure (if none exists the separation question is moot so we make the minimal assumption such a scheme exists) and then modifying it to a scheme that retains MIS-R security but is not DS-secure.

**Proposition 4.11 [MIS-R $\not\rightarrow$ DS]** *Suppose $0 \leq p < 1/2$. Let $\overline{\mathcal{E}}' = \{\mathcal{E}'_k\}_{k\in\mathbb{N}}$ be an encryption scheme with message length $m$ satisfying $m(k) \geq k$ for all $k \in \mathbb{N}$ and with sender ciphertext length $c'$. Assume it is (strongly) MIS-R-secure relative to $\{\mathsf{BSC}_p^{c'(k)}\}_{k\in\mathbb{N}}$. Then there is an encryption scheme $\overline{\mathcal{E}} = \{\mathcal{E}_k\}_{k\in\mathbb{N}}$ with message length $m$ and sender ciphertext length $c$ that is (respectively, strongly) MIS-R-secure relative to $\{\mathsf{BSC}_p^{c(k)}\}_{k\in\mathbb{N}}$ but not (respectively, strongly) DS-secure relative to $\{\mathsf{BSC}_p^{c(k)}\}_{k\in\mathbb{N}}$. Furthermore, if $\overline{\mathcal{E}}'$ is polynomial-time computable so is $\overline{\mathcal{E}}$ and if $\overline{\mathcal{E}}$ is (strongly) decryptable relative to $\{\mathsf{BSC}_p^{c'(k)}\}_{k\in\mathbb{N}}$ then $\overline{\mathcal{E}}$ is (respectively, strongly) decryptable relative to $\{\mathsf{BSC}_p^{c(k)}\}_{k\in\mathbb{N}}$.* ▮

The final conditions regarding computability and decryptability are to ensure that we don't "cheat" by making $\overline{\mathcal{E}}$ different from $\overline{\mathcal{E}}'$ in some categorical way.

**Proof of Proposition 4.11:** Let $n$ be an integer satisfying $e^{-n(0.5-p)^2/2} < 1/4$. Let $c(k) = n + c'(k)$. Let $\mathcal{E}_k(M)$ be defined by

> If $(M = 0^{m(k)}$ OR $M = 1^{m(k)})$ then $a \leftarrow M[1]$ else $a \leftarrow_{\$} \{0,1\}$
> $C' \leftarrow_{\$} \mathcal{E}'_k(M)$
> Return $a^n \| C'$

Recall $b^n$ denotes $n$ copies of the bit $b$. The choice of $n$ ensures that the probability that $\mathsf{BSC}_p^n(a^n)$ has at least $n/2$ positions equal to $a$ is at least $3/4$ for any $a \in \{0,1\}$. (Thus, $a^n$ is a repetition-code based encoding of $a$.) This allows a ds attack with messages $M_0 = 0^{m(k)}$ and $M_1 = 1^{m(k)}$ so that $\mathbf{Adv}^{\mathrm{ds}}(\mathcal{E}_k; \mathsf{BSC}_p^{c(k)}) \geq 1/2$. This implies $\overline{\mathcal{E}}$ is certainly not DS-secure, let alone strongly DS-secure, regardless of the security of $\overline{\mathcal{E}}'$.

Now we want to show that $\overline{\mathcal{E}}$ retains the assumed MIS-R security of $\overline{\mathcal{E}}'$. The reason is that, if the message is random, then the assumption $m(k) \geq k$ implies that it is very unlikely to be either $0^{m(k)}$ or $1^{m(k)}$. But if $M$ is not one of these messages then $\mathcal{E}_k(M)$ provides no more information about $M$ than $\mathcal{E}'_k(M)$. We can thus bound the increase in mis-r-advantage.

Finally, polynomial-time computability is preserved by construction and decryptability because decryption in the new scheme can be done by ignoring the first $n$ bits of the receiver ciphertext and decrypting the remainder as per the old scheme. ▮

### 4.6 Settings in which MIS-R implies MIS

Above we saw that in general MIS-R does not imply MIS. Here we show that for certain types of encryption schemes and channels, MIS-R does imply MIS. This is exploited in [29, 28].

Throughout this section, it is convenient to think of any randomized encryption function $\mathcal{E} : \{0,1\}^m \to \{0,1\}^c$ as a (deterministic) function $\{0,1\}^r \times \{0,1\}^m \to \{0,1\}^c$, where the first argument takes the role of the random coins. We call $\mathcal{E}$ *separable* if

$$\mathcal{E}(R, M) = \mathcal{E}(R, 0^m) \oplus \mathcal{E}(0^r, M) \tag{9}$$

for all $R \in \{0,1\}^r$ and $M \in \{0,1\}^m$. Also, $\mathcal{E}$ is *message linear* if the map $\mathcal{E}(0^r, \cdot) : \{0,1\}^m \to \{0,1\}^c$ is linear, i.e.,

$$\mathcal{E}(0^r, M + M') = \mathcal{E}(0^r, M) + \mathcal{E}(0^r, M') \tag{10}$$

for all $M, M' \in \{0,1\}^m$.

The following theorem states that for an encryption function which is message linear and separable, MIS-R security implies MIS security when the adversarial channel is symmetric and operates independently on ciphertext bits.

**Theorem 4.12 [MIS-R → MIS]** *Let $\mathcal{E} : \{0,1\}^m \to \{0,1\}^c$ be a separable and message-linear encryption function, and let $\mathsf{ChA} : \{0,1\} \to \{0,1\}^\ell$ be a symmetric channel. Then,*

$$\mathbf{Adv}^{\text{mis}}(\mathcal{E}; \mathsf{ChA}^c) \le \mathbf{Adv}^{\text{mis-r}}(\mathcal{E}; \mathsf{ChA}^c) \ . ▮$$

Before we turn to the the proof of Theorem 4.12, we first note that the combination of the encryption function $\mathcal{E} : \{0,1\}^m \to \{0,1\}^c$ and a channel $\mathsf{ChA}^c$ implicitly yields a new channel $\mathsf{Ch}_\mathcal{E} : \{0,1\}^m \to \{0,1\}^{\ell \cdot c}$ which, on input $M \in \{0,1\}^m$ outputs $\mathsf{ChA}^c(\mathcal{E}(M))$. Its output corresponds to the adversarial view. The proof of Theorem 4.12 relies on the channel $\mathsf{Ch}_\mathcal{E}$ being symmetric. In particular, we need the following characterization of symmetric channels.

**Lemma 4.13** *Let $\mathsf{Ch} : \{0,1\}^m \to \{0,1\}^\ell$ be a channel. Assume that there exists a family $\{\tau_M\}_{M \in \{0,1\}^m}$ of functions $\tau_M : \{0,1\}^\ell \to \{0,1\}^\ell$ such that $\tau_{0^m}$ is the identity on $\{0,1\}^\ell$, and moreover, for all $M, M' \in \{0,1\}^m$, and $Y \in \{0,1\}^\ell$,*

$$\tau_{M \oplus M'}(Y) = \tau_M(\tau_{M'}(Y)) \ , \ \ W[M \oplus M', Y] = W[M, \tau_{M'}(Y)] \ . \tag{11}$$

*Then, $\mathsf{Ch}$ is symmetric.* ▮

**Proof of Lemma 4.13:** For every fixed $M \in \{0,1\}^m$, we observe that $\tau_M$ is self-inverse, since $\tau_M(\tau_M(Y)) = \tau_{M \oplus M}(Y) = \tau_{0^m}(Y) = Y$, and hence is a permutation on $\{0,1\}^\ell$. Define the relation $\sim$ such that $Y \sim Y'$ if and only if there exists $M \in \{0,1\}^m$ such that $\tau_M(Y) = Y'$. It is easy to verify that $\sim$ is an equivalence relation.

Let us now partition the columns of the transition matrix $W$ so that two columns corresponding to outputs $Y, Y' \in \{0,1\}^\ell$ are in the same sub-matrix if and only if $Y \sim Y'$. The induced sub-matrices are strongly symmetric: On the one hand, for any two fixed $M, M' \in \{0,1\}^m$, $W[M, Y] = W[M', \tau_{M \oplus M'}(Y)]$ for all $Y \in \{0,1\}^\ell$, and clearly $\tau_{M \oplus M'}(Y) \sim Y$. On the other hand, for any fixed $Y \sim Y'$ there exists $M^* \in \{0,1\}^{\cdot}$ such that $Y' = \tau_{M^*}(Y)$ and thus $W[M, Y] = W[M \oplus M^*, Y']$ for all $M \in \{0,1\}^m$. ▮

**Proof of Theorem 4.12:** We now provide a family of functions $\{\tau_M\}_{M \in \{0,1\}^m}$ as required in Lemma 4.13 for the channel $\mathsf{Ch}_\mathcal{E} : \{0,1\}^m \to \{0,1\}^{\ell \cdot c}$. As $\mathsf{ChA} : \{0,1\} \to \{0,1\}^\ell$ is symmetric, its transition matrix can be decomposed into sub-matrices with the property that, for each sub-matrix, there exist values $p, q \in [0, 1]$ such that all columns are of the form $[p, q]^T$ or $[q, p]^T$, and there is an equal number of columns of each of the two types. Hence, there exists a permutation $\pi_1 : \{0,1\}^\ell \to \{0,1\}^\ell$ such that $\Pr[\,\mathsf{ChA}(1) = y\,] = \Pr[\,\mathsf{ChA}(0) = \pi_1(y)\,]$, and, moreover, $\pi_1^{-1} = \pi_1$. Therefore, with $\pi_0$ being the identity, we have $\Pr[\,\mathsf{ChA}(b) = y\,] = \Pr[\,\mathsf{ChA}(0) = \pi_b(y)\,]$. Also, it is easy to verify that $\pi_{b \oplus b'} = \pi_b \circ \pi_{b'}$ for all $b, b' \in \{0,1\}$. In the same way, when considering $c$-bit inputs, we define a permutation $\pi_X : \{0,1\}^{\ell \cdot c} \to \{0,1\}^{\ell \cdot c}$ such that

$$\pi_X(Y) = (\pi_{X[1]}(Y[1]), \ldots, \pi_{X[c]}(Y[c]))$$

for all $X \in \{0,1\}^c$. Then,

$$\begin{aligned}\Pr[\,\mathsf{ChA}^c(X) = Y\,] &= \prod_{i=1}^{c} \Pr[\,\mathsf{ChA}(X[i]) = Y[i]\,] \\ &= \prod_{i=1}^{c} \Pr\big[\,\mathsf{ChA}(0) = \pi_{X[i]}(Y[i])\,\big] = \Pr[\,\mathsf{ChA}^c(0^c) = \pi_X(Y)\,]\end{aligned}$$

for all $X \in \{0,1\}^c$ and $Y \in \{0,1\}^{\ell \cdot c}$. Consequently, $\pi_{X \oplus X'} = \pi_X \circ \pi_{X'}$ for all $X, X' \in \{0,1\}^c$. We now define, for all $M \in \{0,1\}^m$ and $Y \in \{0,1\}^{\ell \cdot c}$,

$$\tau_M(Y) = \pi_{\mathcal{E}(0^r, M)}(Y)\ .$$

In addition to $\tau_{0^m}$ being the identity (since $\mathcal{E}(0^r, 0^m) = 0^c$), using message linearity we verify that, for all $M, M' \in \{0,1\}^m$,

$$\tau_{M \oplus M'} = \pi_{\mathcal{E}(0^r, M \oplus M')} = \pi_{\mathcal{E}(0^r, M) \oplus \mathcal{E}(0^r, M')} = \pi_{\mathcal{E}(0^r, M)} \circ \pi_{\mathcal{E}(0^r, M')} = \tau_M \circ \tau_{M'}\ .$$

To conclude, with $\mathsf{R}$ being the $r$-bit randomness used by the encryption $\mathcal{E}$,

$$\begin{aligned}\Pr\big[\,\mathsf{Ch}_{\mathcal{E}}(M \oplus M') = Y\,\big] &= \sum_{R \in \{0,1\}^r} \Pr[\,\mathsf{R} = R\,] \cdot \Pr\big[\,\mathsf{ChA}^c(\mathcal{E}(R, M \oplus M')) = Y\,\big] \\ &= \sum_{R \in \{0,1\}^r} \Pr[\,\mathsf{R} = R\,] \cdot \Pr\big[\,\mathsf{ChA}^c(0^c) = \pi_{\mathcal{E}(R, M \oplus M')}(Y)\,\big] \\ &= \sum_{R \in \{0,1\}^r} \Pr[\,\mathsf{R} = R\,] \cdot \Pr\big[\,\mathsf{ChA}^c(0^c) = \pi_{\mathcal{E}(R, M) \oplus \mathcal{E}(0^r, M')}(Y)\,\big] \\ &= \sum_{R \in \{0,1\}^r} \Pr[\,\mathsf{R} = R\,] \cdot \Pr\big[\,\mathsf{ChA}^c(0^c) = \pi_{\mathcal{E}(R, M)}(\pi_{\mathcal{E}(0^r, M')}(Y))\,\big] \\ &= \sum_{R \in \{0,1\}^r} \Pr[\,\mathsf{R} = R\,] \cdot \Pr\big[\,\mathsf{ChA}^c(\mathcal{E}(R, M)) = \pi_{\mathcal{E}(0^r, M')}(Y)\,\big] \\ &= \Pr[\,\mathsf{Ch}_{\mathcal{E}}(M) = \tau_{M'}(Y)\,]\ .\end{aligned}$$

Therefore, $\mathsf{Ch}_{\mathcal{E}}$ is symmetric by Lemma 4.13.

To conclude the proof, we observe that for every symmetric channel $\mathsf{Ch}$ with $m$-bit inputs and every $m$-bit random variable $\mathsf{M}$,

$$\mathbf{I}(\mathsf{M}; \mathsf{Ch}(\mathsf{M})) \le \mathbf{I}(\mathsf{U}; \mathsf{Ch}(\mathsf{U}))$$

where $\mathsf{U}$ is a uniformly distributed $m$-bit string (see e.g. [11, Theorem 7.2.1] for a proof). This implies the theorem statement in the case where $\mathsf{Ch} = \mathsf{Ch}_{\mathcal{E}}$. ▌

Extensions. We do not know how to extend the above result to arbitrary symmetric channels with multi-bit inputs, yet there are several natural channels $\mathsf{ChA}$ which for which MIS-R-security implies MIS-security.

An (important) example is the channel $\mathsf{ChA} : \{0,1\}^c \to \{0,1\}^c$ which, on input $C \in \{0,1\}^c$, samples a fresh noise string $E \in \{0,1\}^c$ according to some given noise distribution, and finally outputs $C \oplus E$. Provided $\mathcal{E} : \{0,1\}^r \times \{0,1\}^m \to \{0,1\}^c$ is separable and message linear, the combined channel $\mathsf{Ch}_{\mathcal{E}}$ is proven to be symmetric by defining $\tau_M$ such that $\tau_M(Y) = \mathcal{E}(0^r, M) \oplus Y$ for all $M \in \{0,1\}^m$ and $Y \in \{0,1\}^c$.

# 5 Achieving DS security

The broad question with regard to achieving security is the following. Let us fix a metric xs, a family $\overline{\mathsf{ChR}}$ of receiver channels and a family $\overline{\mathsf{ChA}}$ of adversary channels. Is there an encryption scheme that is decryptable relative to $\overline{\mathsf{ChR}}$ while achieving XS-security relative to $\overline{\mathsf{ChA}}$?

This question has been examined in the I&C community when XS=MIS-R. The question of interest has been to achieve (and determine) the optimal rate. Results tend to be non-constructive, proving the existence of schemes but not providing explicit schemes, let alone ones that have polynomial time encryption and decryption.

By introducing more demanding security metrics, we have upped the ante. We ask about the achievability of DS (equivalently, SS, MIS). With a practical perspective, we seek not mere existence results but explicit schemes with polynomial-time encryption and decryption. In this section we present such schemes.

The rate of our schemes, although reasonable, is short of optimal. Subsequent work has tackled the fundamental question of determining and achieving the optimal rate, showing, for a wide class of receiver and adversary channels, that the optimal rate for DS security equals the MIS-R secrecy capacity (meaning, the optimal rate for MIS-R security), and presenting schemes that achieve this while having polynomial-time encryption and decryption [3].

The methods we use in this section are based on extractors. (More precisely, what in the cryptographic literature are called strong randomness extractors.) For concreteness we use extractors defined via universal hash functions and their analysis via the Leftover Hash Lemma [20] and its generalizations [16, 15].

Direct use of extractors would, however provide security without error correction, yielding DS-secure schemes when the receiver channel is the clear one. Adding error-correction is not as simple as putting an ECC on top of an encryption that is secure when the receiver channel is the clear one because the ECC helps the adversary. Our approach here is to reduce DS-security to a weaker security requirement that we call rs-r security on an ECC. This is an adaption of the ideas of secure sketches and fuzzy extractors from [16, 15].

## 5.1 Hash functions as extractors

A *hash function* is simply a two-argument function $\mathcal{H}\colon \{0,1\}^h \times \{0,1\}^u \to \{0,1\}^m$. To every "name" or description $h \in \{0,1\}^h$, we associate the function $\mathcal{H}_H = \mathcal{H}(H,\cdot)\colon \{0,1\}^u \to \{0,1\}^m$, meaning $\mathcal{H}_H(U) = \mathcal{H}(H,U)$ for all $U \in \{0,1\}^u$. We say that $\mathcal{H}$ is universal if for all distinct $U_1, U_2 \in \{0,1\}^u$ we have

$$\Pr\left[\,\mathcal{H}(H,U_1) = \mathcal{H}(H,U_2)\,\right] \;\le\; 2^{-m}\ ,$$

where the probability is (only) over $H \leftarrow_{\$} \{0,1\}^h$. We now give some concrete constructions.

The matrix-based construction $\mathcal{H}\colon \{0,1\}^h \times \{0,1\}^u \to \{0,1\}^m$ has $h = um$ and views a description $H \in \{0,1\}^h$ as specifying a $m$ by $u$ matrix over $\mathsf{GF}_2$. It then lets $\mathcal{H}(H,U) = HU$, where $U$ is viewed as a $u$ by 1 matrix over $\mathsf{GF}_2$ and $HU$ is matrix-vector multiplication, returning a 1 by $m$ matrix over $\mathsf{GF}_2$ that is regarded as an $m$-bit string.

When $m \le u$, a construction reducing the description length $h$ from $um$ to just $u$ can be obtained as follows. Identify $\{0,1\}^\ell$ with $\mathsf{GF}_{2^\ell}$ for any $\ell$. Fix a regular projection $\pi\colon \mathsf{GF}_{2^u} \to \mathsf{GF}_{2^m}$, meaning that every $y \in \mathsf{GF}_{2^m}$ has exactly $2^{u-m}$ pre-images under $\pi$. The description $H$ of a function is a point $H \in \mathsf{GF}_{2^u}$, and we let $\mathcal{H}(H,U) = \pi(HU)$. Here the multiplication is in $\mathsf{GF}_{2^u}$. These are all standard constructions whose universality is easily checked [43, Chapter 8].

The following generalization of the Leftover Hash Lemma of [20] was stated in [16, 15].

| **Transform** $\mathcal{E}(M)$ | **Transform** $\mathcal{D}(Y)$ |
|---|---|
| $H \leftarrow_\$ \{0,1\}^h$ ; $U \leftarrow_\$ \{0,1\}^u$ | $Y_1 \| Y_2 \leftarrow C$ |
| $P \leftarrow \mathcal{H}(H,U)$ ; $W \leftarrow P \oplus M$ | $U \leftarrow \mathsf{En1}^{-1}(Y_1)$ ; $H \| W \leftarrow \mathsf{En2}^{-1}(Y_2)$ |
| $X \leftarrow \mathsf{En1}(U) \| \mathsf{En2}(H \| W)$ | $P \leftarrow \mathcal{H}(H,U)$ ; $M \leftarrow P \oplus W$ |
| Return $X$ | Return $M$ |

Figure 3: On the left is the encryption function $\mathcal{E} = \mathbf{XtX}[\mathcal{H}, \mathsf{En1}, \mathsf{En2}]\colon \{0,1\}^m \to \{0,1\}^{n_1+n_2}$ associated to hash function $\mathcal{H}\colon \{0,1\}^h \times \{0,1\}^u \to \{0,1\}^m$ and ECCs $\mathsf{En1}\colon \{0,1\}^u \to \{0,1\}^{n_1}$ and $\mathsf{En2}\colon \{0,1\}^{h+m} \to \{0,1\}^{n_2}$ by the **XtX** construction. On the right is the decryption function for a channel $\mathsf{ChR1} \| \mathsf{ChR2}$ for which $\mathsf{En1}, \mathsf{En2}$ are, respectively, ECCs.

---

**Lemma 5.1 [Generalized Leftover Hash Lemma]** *Let $\mathcal{H}\colon \{0,1\}^h \times \{0,1\}^u \to \{0,1\}^m$ be a universal hash function. Let $\mathsf{U}$ be a random variable over $\{0,1\}^u$. Let random variable $\mathsf{H}$ be uniformly distributed over $\{0,1\}^h$. Let random variable $\mathsf{V}$ be uniformly distributed over $\{0,1\}^m$. Let $\mathsf{Z}$ be a random variable. Assume $(\mathsf{U}, \mathsf{Z}), \mathsf{H}, \mathsf{V}$ are independent. Then*

$$\mathbf{SD}((\mathsf{H}, \mathsf{Z}, \mathcal{H}(\mathsf{H}, \mathsf{U})); (\mathsf{H}, \mathsf{Z}, \mathsf{V})) \;\leq\; \frac{1}{2}\sqrt{2^m \cdot \mathbf{GP}(\mathsf{U}|\mathsf{Z})}\;. \;\blacksquare$$

Above, $\mathsf{Z}$ may depend on $\mathsf{U}$ but the three random variables $(\mathsf{U}, \mathsf{Z}), \mathsf{H}, \mathsf{V}$ must be independent.

### 5.2 The XtX construction

Let $\mathcal{H}\colon \{0,1\}^h \times \{0,1\}^u \to \{0,1\}^m$ be a hash function. Let $\mathsf{En1}\colon \{0,1\}^u \to \{0,1\}^{n_1}$ and $\mathsf{En2}\colon \{0,1\}^{h+m} \to \{0,1\}^{n_2}$ be injective functions. These will later be instantiated by (the encoding functions of) ECCs. The **XtX** (e<u>x</u>tract <u>t</u>hen <u>x</u>or) construction associates to $\mathcal{H}, \mathsf{En1}, \mathsf{En2}$ the encryption function $\mathcal{E} = \mathbf{XtX}[\mathcal{H}, \mathsf{En1}, \mathsf{En2}]\colon \{0,1\}^m \to \{0,1\}^{n_1+n_2}$ defined via Figure 3. The encryption function picks at random a $h$-bit string $H$ to specify hash function $\mathcal{H}(H, \cdot)$ as well as a random $r$-bit string $U$, and hashes $U$ under $\mathcal{H}(H, \cdot)$ to obtain a $m$-bit pad $P$. The latter is xored to the message $M \in \{0,1\}^m$ to get $W$. The sender ciphertext $X$ includes $W$, but also includes $H, X$ to enable decryption.

### 5.3 Overview

We will analyze **XtX** in a general setting. The only assumption made about the channel (whether receiver or adversary) $\mathsf{Ch}\colon \{0,1\}^{n_1+n_2} \to \{0,1\}^d$ is that it is $(n_1, n_2)$-splittable, meaning the applications of the channel on $\mathsf{En1}(U)$ and on $\mathsf{En2}(H \| W)$ are independent. The canonical setting is that both the receiver and adversary channel are induced by binary channels, in which case they are certainly $(n_1, n_2)$-splittable. The binary channels here may be symmetric but need not be. Within this general setting, we look at decryptability and DS-security and establish the following:

- Say the receiver channel splits as $\mathsf{ChR} = \mathsf{ChR1} \| \mathsf{ChR2}$. If $\mathsf{En1}$ is a good ECC for $\mathsf{ChR1}$ and $\mathsf{En2}$ is a good ECC for $\mathsf{ChR2}$ then $\mathbf{XtX}[\mathcal{H}, \mathsf{En1}, \mathsf{En2}]$ is decryptable. (By "good" here we mean that decoding is possible with low error.)
- Say the adversary channel splits as $\mathsf{ChA} = \mathsf{ChA1} \| \mathsf{ChA2}$. DS-security of $\mathbf{XtX}[\mathcal{H}, \mathsf{En1}, \mathsf{En2}]$ relative to $\mathsf{ChA}$ makes no assumptions about $\mathsf{ChR2}$ or $\mathsf{En2}$, meaning holds for *all* choices of these. The requirements will be that $\mathcal{H}$ is universal and that $\mathsf{En1}$, viewed as an encryption function, provides a certain weak security property, called rs-r (recovery security for random messages), relative to $\mathsf{ChA1}$. This asks that if $\mathsf{U}$ is uniformly distributed then it is hard to recover it from $\mathsf{ChA1}(\mathsf{En1}(\mathsf{U}))$. This notion and approach are inspired by the secure sketches of [16, 15].

Now, given a particular choice of receiver and adversary channels, we would pick $\mathsf{En1}, \mathsf{En2}$ to give us the error-correction and security properties needed. To exemplify we will consider the case where $\mathsf{ChR}, \mathsf{ChA}$ are induced by binary channels and in particular the binary symmetric channel. For these cases we will provide bounds on the rs-r advantage of arbitrary ECCs. With this in hand, concrete solutions emerge for these channels.

Both encryption and decryption will be polynomial-time assuming the codes have polynomial-time encoding and decoding. The rate of $\mathbf{XtX}[\mathcal{H}, \mathsf{En1}, \mathsf{En2}]$ depends on the choice of codes but even with optimal choices is not itself optimal. It turns out that $H$, once chosen and transmitted, can be re-used across multiple encryptions, playing the role of a seed in what in [3] is called a seeded-encryption scheme, so, via amortization, it can be ignored from the point of view of rate. This increases the rate but still leaves it short of optimal. We are not aware of any simple way to fill gap. A scheme with optimal rate is given in [3] using alternative techniques.

### 5.4 Decryptability of XtX

Consider any receiver channel $\mathsf{ChR}$ that has the form $\mathsf{ChR} = \mathsf{ChR1} \| \mathsf{ChR2}$ where $\mathsf{ChR1}\colon \{0,1\}^{n_1} \to \{0,1\}^{d_1}$ and $\mathsf{ChR2}\colon \{0,1\}^{n_2} \to \{0,1\}^{d_2}$. If $\mathsf{En1}, \mathsf{En2}$ are, respectively, good ECCs over $\mathsf{ChR1}, \mathsf{ChR2}$, then $\mathbf{XtX}[\mathcal{H}, \mathsf{En1}, \mathsf{En2}]$ will be decryptable over $\mathsf{ChR}$. Decryption is performed as shown in Figure 3. The first step parses receiver ciphertext $Y$ into its first $d_1$ bits $Y_1$ and last $d_2$ bits $Y_2$. Decryption (ie. decoding) algorithms for the ECCs are then applied, and the outputs again appropriately parsed.

**Theorem 5.2 [Decryptability of XtX]** *Let $\mathcal{H}\colon \{0,1\}^h \times \{0,1\}^u \to \{0,1\}^m$ be a hash function. Let $\mathsf{En1}^{-1}\colon \{0,1\}^{d_1} \to \{0,1\}^r$ be a decryption function for $\mathsf{En1}\colon \{0,1\}^u \to \{0,1\}^{n_1}$ over channel $\mathsf{ChR1}\colon \{0,1\}^{n_1} \to \{0,1\}^{d_1}$ with decryption error $\delta_1$. Let $\mathsf{En2}^{-1}\colon \{0,1\}^{d_2} \to \{0,1\}^{k+m}$ be a decryption function for $\mathsf{En2}\colon \{0,1\}^{h+m} \to \{0,1\}^{n_2}$ over channel $\mathsf{ChR2}\colon \{0,1\}^{n_2} \to \{0,1\}^{d_2}$ with decryption error $\delta_2$. Let $\mathcal{E} = \mathbf{XtX}[\mathcal{H}, \mathsf{En1}, \mathsf{En2}]$. Then $\mathcal{D}\colon \{0,1\}^{d_1+d_2} \to \{0,1\}^m$ as depicted in Figure 3 is a decryption function for $\mathcal{E}$ with decryption error at most $\delta_1 + \delta_2$.* ▮

### 5.5 DS-security of XtX

Regard $\mathsf{En1}\colon \{0,1\}^u \to \{0,1\}^{n_1}$ as an encryption function. We define recovery security for random messages (rs-r) relative to channel $\mathsf{ChA1}\colon \{0,1\}^{n_1} \to \{0,1\}^{d_1}$ via the advantage

$$\mathbf{Adv}^{\text{rs-r}}(\mathsf{En1}; \mathsf{ChA1}) \;=\; \mathbf{GP}(\mathsf{U} | \mathsf{ChA1}(\mathsf{En1}(\mathsf{U}))) \;=\; \max_{\mathcal{A}} \Pr[\mathcal{A}(\mathsf{ChA1}(\mathsf{En1}(\mathsf{U}))) = \mathsf{U}]$$

where random variable $\mathsf{U}$ is uniformly distributed over $\{0,1\}^u$ and the maximum is over all adversaries $\mathcal{A}$. As this shows the definition can either be expressed in terms of information-theoretic quantities (the guessing probability) or using adversaries, and we will have occasion to exploit both representations.

**Theorem 5.3 [DS-security of XtX]** *Let $\mathcal{H}\colon \{0,1\}^h \times \{0,1\}^u \to \{0,1\}^m$ be a universal hash function. Let $\mathsf{En1}\colon \{0,1\}^u \to \{0,1\}^{n_1}$ and $\mathsf{En2}\colon \{0,1\}^{h+m} \to \{0,1\}^{n_2}$ be injective functions. Let $\mathcal{E} = \mathbf{XtX}[\mathcal{H}, \mathsf{En1}, \mathsf{En2}]$. Let $\mathsf{ChA1}\colon \{0,1\}^{n_1} \to \{0,1\}^{d_1}$ and $\mathsf{ChA2}\colon \{0,1\}^{n_2} \to \{0,1\}^{n_2}$ be channels. Let $\mathsf{ChA} = \mathsf{ChA1} \| \mathsf{ChA2}$. Then*

$$\mathbf{Adv}^{\text{ds}}(\mathcal{E}; \mathsf{ChA}) \;\leq\; \sqrt{2^m \cdot \mathbf{Adv}^{\text{rs-r}}(\mathsf{En1}; \mathsf{ChA1})}\;. \;▮$$

The bound does not depend on $\mathsf{ChA2}, \mathsf{En2}$, reflecting that security is independent of the choice of these. The proof will use the following standard lemma.

**Lemma 5.4** *Let $\mathsf{X}_1, \mathsf{X}_2$ be random variables and let $f_1, f_2$ be transforms. Then $\mathbf{SD}(f_1(\mathsf{X}_1); f_2(\mathsf{X}_2)) \leq \mathbf{SD}(\mathsf{X}_1; \mathsf{X}_2)$.* ▮

Given this we proceed to prove Theorem 5.3.

**Proof of Theorem 5.3:** Let $M_0, M_1 \in \{0,1\}^m$. Let $\mathsf{U}, \mathsf{H}, \mathsf{V}$ be uniformly and independently distributed random variables, the first over $\{0,1\}^u$, the second over $\{0,1\}^h$ and the third over $\{0,1\}^m$. Let $\mathsf{Z} = \mathsf{ChA1}(\mathsf{En1}(\mathsf{U}))$. For $c \in \{0,1\}$ let $\mathsf{W}_c = \mathsf{ChA2}(\mathsf{En2}(\mathsf{H} \| \mathcal{H}(\mathsf{H}, \mathsf{U}) \oplus M_c))$. Let

$$\begin{aligned} \delta(M_0, M_1) &= \mathbf{SD}((\mathsf{Z}, \mathsf{W}_0); (\mathsf{Z}, \mathsf{W}_1)) \\ \epsilon(M_0, M_1) &= \mathbf{SD}((\mathsf{H}, \mathsf{Z}, \mathcal{H}(\mathsf{H}, \mathsf{U}) \oplus M_0); (\mathsf{H}, \mathsf{Z}, \mathcal{H}(\mathsf{H}, \mathsf{U}) \oplus M_1))\;. \end{aligned}$$

Lemma 5.4 implies that regardless of the choices of $\mathsf{En2}$ and $\mathsf{ChA2}$ we have

$$\delta(M_0, M_1) \;\le\; \epsilon(M_0, M_1)\;.$$

We will now show that

$$\epsilon(M_0, M_1) \;\le\; \sqrt{2^m \cdot \mathbf{Adv}^{\text{rs-r}}(\mathsf{En1}; \mathsf{ChA1})}\;. \tag{12}$$

Since $M_0, M_1$ were arbitrary, this proves the theorem.

Towards proving (12), let

$$\begin{aligned}
\epsilon_1(M_0, M_1) &= \mathbf{SD}((\mathsf{H}, \mathsf{Z}, \mathcal{H}(\mathsf{H}, \mathsf{U}) \oplus M_0); (\mathsf{H}, \mathsf{Z}, \mathsf{V} \oplus M_0))\\
\epsilon_2(M_0, M_1) &= \mathbf{SD}((\mathsf{H}, \mathsf{Z}, \mathsf{V} \oplus M_0), (\mathsf{H}, \mathsf{Z}, \mathsf{V} \oplus M_1))\\
\epsilon_3(M_0, M_1) &= \mathbf{SD}((\mathsf{H}, \mathsf{Z}, \mathsf{V} \oplus M_1), (\mathsf{H}, \mathsf{Z}, \mathcal{H}(\mathsf{H}, \mathsf{U}) \oplus M_1))\;.
\end{aligned}$$

By the triangle inequality we have

$$\epsilon(M_0, M_1) \;\le\; \epsilon_1(M_0, M_1) + \epsilon_2(M_0, M_1) + \epsilon_3(M_0, M_1)\;.$$

However, $\epsilon_2(M_0, M_1) = 0$ because $\mathsf{V}$ is uniformly distributed over $\{0,1\}^m$ and $M_0, M_1 \in \{0,1\}^m$. Also,

$$\epsilon_1(M_0, M_1) \;=\; \epsilon_3(M_0, M_1) \;=\; \mathbf{SD}((\mathsf{H}, \mathsf{Z}, \mathcal{H}(\mathsf{H}, \mathsf{U})), (\mathsf{H}, \mathsf{Z}, \mathsf{V}))\;.$$

We conclude by applying Lemma 5.1. ▮

## 5.6 XtX over a BSC

We can apply the above general framework and results to provide concrete DS-secure encryption schemes for particular channels. We illustrate here for the case of the binary symmetric channel. We begin for simplicity with the case that the receiver channel is error-free and then move on to the case where it is not.

Assume the receiver channel is error-free. In this case we can set $\mathsf{En1} = \mathsf{Id}_u$ and $\mathsf{En2} = \mathsf{Id}_{h+m}$ to identity functions. Rs-r advantages are now easy to bound. In the case of the binary symmetric channel we have the following:

**Proposition 5.5** *Suppose* $0 \le p \le 1/2$ *and let* $\mathsf{ChA} = \mathsf{BSC}_p^u$. *Then*

$$\mathbf{Adv}^{\text{rs-r}}(\mathsf{Id}_u; \mathsf{ChA}) \;=\; (1-p)^u\;. \;▮ \tag{13}$$

**Proof:** Letting random variable $\mathsf{U}$ be uniformly distributed over $\{0,1\}^u$ we have

$$\begin{aligned}
\mathbf{GP}(\mathsf{U}|\mathsf{BSC}_p^u(\mathsf{U})) &= \sum_z \Pr[\mathsf{BSC}_p^u(\mathsf{U}) = z] \cdot \max_x \Pr[\mathsf{U} = x | \mathsf{BSC}_p^u(\mathsf{U}) = z]\\
&= \sum_z \Pr[\mathsf{BSC}_p^u(\mathsf{U}) = z] \cdot \Pr[\mathsf{U} = z | \mathsf{BSC}_p^u(\mathsf{U}) = z]\\
&= \sum_z \Pr[\mathsf{BSC}_p^u(\mathsf{U}) = z] \cdot (1-p)^u\\
&= \sum_z 2^{-u} \cdot (1-p)^u\\
&= (1-p)^u\;,
\end{aligned}$$

establishing (13). ▮

By combining Theorem 5.3 and Proposition 5.5 we get the following concrete bound for DS-security of **XtX** encryption in the case the receiver channel is error-free and the adversary channel is a binary symmetric one:

**Theorem 5.6** *Let* $\mathcal{H}\colon \{0,1\}^h \times \{0,1\}^u \to \{0,1\}^m$ *be a universal hash function. let* $\mathcal{E} = \mathbf{XtX}[\mathcal{H}, \mathsf{Id}_u, \mathsf{Id}_{h+m}]$. *Suppose* $0 \le p \le 1/2$ *and let* $\mathsf{ChA} = \mathsf{BSC}_p^{u+h+m}$.

$$\mathbf{Adv}^{\text{ds}}(\mathcal{E}; \mathsf{ChA}) \;\le\; \sqrt{2^m \cdot (1-p)^u}\;. \;▮$$

Now let us consider the case where the receiver channel is not error-free. $\mathsf{En1}, \mathsf{En2}$ would now be chosen to correct errors over the receiver channel. Taking them as given, we need to bound the rs-r advantage of the former. For the case where the adversary channel is a binary symmetric one we have the following:

**Proposition 5.7** *Let* $\mathsf{En1}\colon \{0,1\}^u \to \{0,1\}^{u+r}$ *be an injective function. Suppose* $0 \le p \le 1/2$ *and let* $\mathsf{ChA} = \mathsf{BSC}_p^{u+r}$. *Then*

$$\mathbf{Adv}^{\text{rs-r}}(\mathsf{En1}; \mathsf{ChA}) \;\le\; 2^r(1-p)^{u+r} \; . \; \blacksquare \tag{14}$$

**Proof:** Let $\mathsf{Z} = \mathsf{ChA}(\mathsf{En1}(\mathsf{U})$ where random variable $\mathsf{U}$ is uniformly distributed over $\{0,1\}^u$. Then we have

$$\begin{aligned}
\mathbf{GP}(\mathsf{U}|\mathsf{Z}) &= \sum_z \Pr[\mathsf{Z}=z] \cdot \max_x \Pr[\,\mathsf{U}=x \mid \mathsf{Z}=z\,] \\
&= \sum_z \Pr[\mathsf{Z}=z] \cdot \max_x \Pr[\,\mathsf{Z}=z \mid \mathsf{U}=x\,] \frac{\Pr[\mathsf{U}=x]}{\Pr[\mathsf{Z}=z]} \\
&= \sum_z \max_x \Pr[\,\mathsf{Z}=z \mid \mathsf{U}=x\,] \cdot 2^{-u} \\
&\le \sum_z 2^{-u}(1-p)^{u+r} \\
&= 2^r(1-p)^{u+r} \, ,
\end{aligned}$$

which proves the Proposition. $\blacksquare$

From Theorem 5.3 and the above we directly obtain the following:

**Theorem 5.8** *Let* $\mathcal{H}\colon \{0,1\}^h \times \{0,1\}^u \to \{0,1\}^m$ *be a universal hash function. Let* $\mathsf{En1}\colon \{0,1\}^u \to \{0,1\}^{u+r}$ *and* $\mathsf{En2}\colon \{0,1\}^{h+m} \to \{0,1\}^{n_2}$ *be injective functions. Let* $\mathcal{E} = \mathbf{XtX}[\mathcal{H}, \mathsf{En1}, \mathsf{En2}]$. *Suppose* $0 \le p \le 1/2$ *and let* $\mathsf{ChA} = \mathsf{BSC}_p^{u+r+n_2}$. *Then*

$$\mathbf{Adv}^{\text{ds}}(\mathcal{E}; \mathsf{ChA}) \;\le\; \sqrt{2^{m+r} \cdot (1-p)^{u+r}} \; . \; \blacksquare$$

## 5.7 Numerical estimates

In usage, we would imagine the user or system designer wanting to create a scheme that encrypts messages of a certain desired length $m$ and achieves a certain desired number of bits $s$ of DS-security under the assumption that the adversary channel is a BSC with a certain, given crossover probability $p$. Our job, given $m, s, p$, is to provide the scheme. The above allows us to do so. To illustrate, let us for simplicity consider the case where the receiver channel is error-free, so that $\mathcal{E} = \mathbf{XtX}[\mathcal{H}, \mathsf{Id}_u, \mathsf{Id}_{h+m}]$. Our task amounts to picking $\mathcal{H}$ so that $\mathbf{Adv}^{\text{ds}}(\mathcal{E}; \mathsf{BSC}_p^u) \le 2^{-s}$. Theorem 5.6 says that it suffices to let

$$u \;=\; \left\lceil \frac{2s+m}{\alpha} \right\rceil \qquad \text{where} \qquad \alpha = \lg \frac{1}{1-p} \; .$$

We can then use one of the constructions of $\mathcal{H}\colon \{0,1\}^h \times \{0,1\}^u \to \{0,1\}^m$ given above. The amortized rate of the scheme is

$$\mathbf{Rate}(\mathcal{E}) \;=\; \frac{m}{u+m} \;=\; \frac{\alpha m}{2s+m+\alpha m}$$

which is

$$\overline{\mathbf{Rate}}(\mathcal{E}) \;=\; \frac{\alpha}{1+\alpha}$$

in the limit as $m \to \infty$. By the amortized rate we mean that we ignore $H$ with the understanding that it can be chosen and transmitted just once, all subsequent encryptions using the same $H$, so the amortized rate does not depend on $h$.

| $p$ | 0.5 | 0.4 | 0.3 | 0.2 | 0.1 |
|---|---|---|---|---|---|
| $\overline{\mathbf{Rate}}(\mathcal{E})$ | 0.5 | 0.42 | 0.34 | 0.24 | 0.13 |
| $\overline{\mathbf{Rate}}_2(\mathcal{E})$ | 1 | 0.74 | 0.51 | 0.32 | 0.15 |

Figure 4: For various values of the crossover probability $p$ of the adversary BSC, we show the limiting amortized rate and dual-channel limiting amortized rate achievable by $\mathbf{XtX}[\mathcal{H}, \mathsf{Id}_u, \mathsf{Id}_{h+m}]$.

---

We note another improvement. As we saw above, security is not affected by $\mathsf{ChA2}$, the channel over which $W$ is sent. (We have already sent $H$ upfront, so only $W$ goes over this channel.) The sender could thus transmit $W$ separately over a clear channel. The adversary would be in full possession of $W$ but security is not reduced. The advantage is that transmission in a way that renders the adversary's reception noisy is much more expensive than transmission in the clear, and sending $W$ becomes essentially for free. Let us refer to this as the dual-channel setting and let $\mathbf{Rate}_2$ denote the corresponding rate, which is

$$\mathbf{Rate}_2(\mathcal{E}) \;=\; \frac{m}{u} \;=\; \frac{\alpha m}{2s+m}\;.$$

In the limit as $m \to \infty$ this is

$$\overline{\mathbf{Rate}}_2(\mathcal{E}) \;=\; \alpha\;.$$

In Figure 4 we illustrate, showing, for a few choices of $p$ the corresponding rates. The rate obviously decreases with $p$ but the salient fact is that we can get security for any positive value of $p$, even if the rate is quite low.

We note yet another improvement, namely that the $1-p$ term in Theorems 5.6 and 5.8 can be replaced with $1-2p+p^2$ by using a version of Lemma 5.1 in which the guessing probability is replaced by a collision probability. Our numbers do not reflect this.

### 5.8 Reduction to error-free channels

Above we applied Theorem 5.3 in the case where the adversary channel is the binary symmetric one. Here we look into applying it for other channels.

In general $\mathsf{En1}$ is chosen to correct errors over the receiver channel, and it may be hard to bound its rs-r advantage. But in the case that the receiver channel is error-free, $\mathsf{En1}$ is just the identity function $\mathsf{Id}_u$, and bounding rs-r advantage should be much easier, as we saw above. Here we aim to generalize this approach. For a wide class of adversary channels and choices of $\mathsf{En1}$, we will show how to reduce the problem of bounding the rs-r advantage of $\mathsf{En1}$ over the given channel to the problem of bounding the rs-r advantage of $\mathsf{Id}_u$ over the same channel. This effectively makes security independent of the details of the ECCs. More specifically, what we show is that for systematic codes (most codes are systematic or can be made so), and for appropriately splittable adversary channels (all binary channels fall in this category), the rs-r security of the code over the given channel depends only on the amount of redundancy in the code and the rs-r security of the identity function over the same channel.

Let us now proceed to the details. An ECC $\mathsf{En1}\colon \{0,1\}^u \to \{0,1\}^{n_1}$ is *systematic* if there is a *redundancy function* $\mathsf{Rd}\colon \{0,1\}^u \to \{0,1\}^{n_1-u}$ such that $\mathsf{En1}(U) = U \| \mathsf{Rd}(U)$ for all $U \in \{0,1\}^u$. Then we have the following:

**Theorem 5.9** *Let* $\mathsf{En1}\colon \{0,1\}^u \to \{0,1\}^{u+r}$ *be a systematic ECC with redundancy function* $\mathsf{Rd}\colon \{0,1\}^u \to \{0,1\}^r$. *Let* $\mathsf{Ch}_u\colon \{0,1\}^u \to \{0,1\}^a$ *and* $\mathsf{Ch}_r\colon \{0,1\}^r \to \{0,1\}^b$ *be channels, and let* $\mathsf{ChA1} = \mathsf{Ch}_u \| \mathsf{Ch}_r\colon \{0,1\}^{u+r} \to \{0,1\}^{a+b}$. *Then*

$$\mathbf{Adv}^{\text{rs-r}}(\mathsf{En1}; \mathsf{ChA1}) \;\leq\; 2^r \cdot \mathbf{Adv}^{\text{rs-r}}(\mathsf{Id}_u; \mathsf{Ch}_u)\;. \quad \blacksquare \qquad (15)$$

In the case that ChA is the binary symmetric channel, we could combine Proposition 5.5 and Theorem 5.9 to bound the rs-r advantage of a given En1 assuming it was systematic. Specifically, if En1: $\{0,1\}^u \to \{0,1\}^{u+r}$ is a systematic ECC and $\mathsf{ChA} = \mathsf{BSC}_p^{u+r}$ then we would get $\mathbf{Adv}^{\text{rs-r}}(\mathsf{En1}; \mathsf{ChA}) \le 2^r(1-p)^u$. This is not as good a bound as Proposition 5.7. In the case of BSCs, Theorem 5.9 is thus not as effective as a direct analysis. But it is more general, and for channels other than BSCs it may be hard to directly analyze rs-r security of En1. In this case Theorem 5.9 will be helpful.

To prove Theorem 5.9 we first need a few definitions and lemmas. If $\mathsf{Ch}_1\colon D_1 \to R_1$ and $\mathsf{Ch}_2\colon D_2 \to R_2$ are channels with $R_1 \subseteq D_2$ then their composition $\mathsf{Ch} = \mathsf{Ch}_2 \circ \mathsf{Ch}_1$ is the channel $\mathsf{Ch}\colon D_1 \to R_2$ defined by $\mathsf{Ch}(x) = \mathsf{Ch}_2(\mathsf{Ch}_1(x))$ for all $x \in D_1$. We say that channel $\mathsf{Ch}_3$ is a *degradation* of channel $\mathsf{Ch}_1$ if there is a channel $\mathsf{Ch}_2$ such that $\mathsf{Ch}_3 = \mathsf{Ch}_2 \circ \mathsf{Ch}_1$.

**Lemma 5.10** *Let* En1: $\{0,1\}^u \to \{0,1\}^{n_1}$ *be a function. Let* $\mathsf{Ch}_1\colon \{0,1\}^{n_1} \to \{0,1\}^c$ *and* $\mathsf{Ch}_2\colon \{0,1\}^c \to \{0,1\}^{d_1}$ *be channels and let* $\mathsf{Ch} = \mathsf{Ch}_2 \circ \mathsf{Ch}_1$. *Then*

$$\mathbf{Adv}^{\text{rs-r}}(\mathcal{E}; \mathsf{Ch}) \;\le\; \mathbf{Adv}^{\text{rs-r}}(\mathcal{E}; \mathsf{Ch}_1) \tag{16}$$

**Proof:** Let random variable U be uniformly distributed over $\{0,1\}^u$. It suffices to show that

$$\max_{\mathcal{A}} \Pr[\mathcal{A}(\mathsf{Ch}(\mathsf{En1}(\mathsf{U}))) = \mathsf{U}] \;\le\; \max_{\mathcal{B}} \Pr[\mathcal{B}(\mathsf{Ch}_1(\mathsf{En1}(\mathsf{U}))) = \mathsf{U}]\;. \tag{17}$$

To show this, let $\mathcal{A}$ be any adversary. We define adversary $\mathcal{B}_{\mathcal{A}}$ via $\mathcal{B}_{\mathcal{A}}(Z) = \mathcal{A}(\mathsf{Ch}_2(Z))$ for all $Z$. Then for any $\mathcal{A}$ we have

$$\Pr[\mathcal{A}(\mathsf{Ch}(\mathsf{En1}(\mathsf{U}))) = \mathsf{U}] \;=\; \Pr[\mathcal{B}_{\mathcal{A}}(\mathsf{Ch}_1(\mathsf{En1}(\mathsf{U}))) = \mathsf{U}]\;. \tag{18}$$

Now take the max over all $\mathcal{A}$ of both sides of (18) to get

$$\begin{aligned}
\max_{\mathcal{A}} \Pr[\mathcal{A}(\mathsf{Ch}(\mathsf{En1}(\mathsf{U}))) = \mathsf{U}] &= \max_{\mathcal{A}} \Pr[\mathcal{B}_{\mathcal{A}}(\mathsf{Ch}_1(\mathsf{En1}(\mathsf{U}))) = \mathsf{U}] \\
&\le \max_{\mathcal{B}} \Pr[\mathcal{B}(\mathsf{Ch}_1(\mathsf{En1}(\mathsf{U}))) = \mathsf{U}]\;,
\end{aligned}$$

the last inequality because the max over all $\mathcal{B}$ includes those of the form $\mathcal{B}_{\mathcal{A}}$. ▮

The following lemma about guessing probabilities is a corollary of [15, Lemma 2.2] but for completeness we give a direct proof.

**Lemma 5.11** *Let* X, R, Z *be random variables such that* R *takes on at most* $N \ge 1$ *values. Then*

$$\mathbf{GP}(\mathsf{X}|\mathsf{Z},\mathsf{R}) \;\le\; N \cdot \mathbf{GP}(\mathsf{X}|\mathsf{Z})\;. \;\blacksquare \tag{19}$$

**Proof:** We have

$$\mathbf{GP}(\mathsf{X}|\mathsf{Z},\mathsf{R}) \;=\; \sum_{r,z} \Pr[\mathsf{R} = r, \mathsf{Z} = z] \cdot f(r,z) \tag{20}$$

where

$$\begin{aligned}
f(r,z) &= \max_x \Pr\left[\,\mathsf{X} = x \;|\; \mathsf{R} = r, \mathsf{Z} = z\,\right] \\
&= \max_x \frac{\Pr[\mathsf{X} = x, \mathsf{R} = r, \mathsf{Z} = z]}{\Pr[\mathsf{R} = r, \mathsf{Z} = z]} \\
&= \max_x \frac{\Pr[\mathsf{X} = x, \mathsf{R} = r | \mathsf{Z} = z] \cdot \Pr[\mathsf{Z} = z]}{\Pr[\mathsf{R} = r | \mathsf{Z} = z] \cdot \Pr[\mathsf{Z} = z]} \\
&= \max_x \frac{\Pr[\mathsf{X} = x, \mathsf{R} = r | \mathsf{Z} = z]}{\Pr[\mathsf{R} = r | \mathsf{Z} = z]} \\
&\le \max_x \frac{\Pr[\mathsf{X} = x | \mathsf{Z} = z]}{\Pr[\mathsf{R} = r | \mathsf{Z} = z]}\;.
\end{aligned} \tag{21}$$

From (20) and (21) we have

$$\begin{aligned}
\mathbf{GP}(\mathsf{X}|\mathsf{Z},\mathsf{R}) &\leq \sum_{r,z} \Pr[\mathsf{R}=r,\mathsf{Z}=z]\cdot\max_x \frac{\Pr[\mathsf{X}=x|\mathsf{Z}=z]}{\Pr[\mathsf{R}=r|\mathsf{Z}=z]} \\
&= \sum_{r,z} \Pr[\mathsf{R}=r|\mathsf{Z}=z]\cdot\Pr[\mathsf{Z}=z]\cdot\max_x \frac{\Pr[\mathsf{X}=x|\mathsf{Z}=z]}{\Pr[\mathsf{R}=r|\mathsf{Z}=z]} \\
&= \sum_{r,z} \Pr[\mathsf{Z}=z]\cdot\max_x \Pr[\mathsf{X}=x|\mathsf{Z}=z] \\
&= \sum_{r} \mathbf{GP}(\mathsf{X}|\mathsf{Z}) \\
&\leq N\cdot\mathbf{GP}(\mathsf{X}|\mathsf{Z})\ .
\end{aligned}$$

This concludes the proof. ▮

Now we proceed to prove Theorem 5.9.

**Proof of Theorem 5.9:** $\mathsf{ChA1} = \mathsf{Ch}_u\|\mathsf{Ch}_r$ is a degradation of $\mathsf{Ch}_u\|\mathsf{Id}_r$ hence by Lemma 5.10 we have

$$\mathbf{Adv}^{\text{rs-r}}(\mathsf{En1};\mathsf{Ch}) \leq \mathbf{Adv}^{\text{rs-r}}(\mathsf{En1};\mathsf{Ch}_u\|\mathsf{Id}_r)\ .$$

We proceed to upper bound the latter and thereby establish (15) as follows:

$$\begin{aligned}
\mathbf{Adv}^{\text{rs-r}}(\mathsf{En1};\mathsf{Ch}_u\|\mathsf{Id}_r) &= \mathbf{GP}(\mathsf{U}|\mathsf{Ch}_u(\mathsf{U}),\mathsf{Rd}(\mathsf{U})) && (22) \\
&\leq 2^r\cdot\mathbf{GP}(\mathsf{U}|\mathsf{Ch}_u(\mathsf{U})) && (23) \\
&= 2^r\cdot\mathbf{Adv}^{\text{rs-r}}(\mathsf{Id}_u;\mathsf{Ch}_u)\ .
\end{aligned}$$

Since $\mathsf{En1}$ is systematic with redundancy function $\mathsf{Rd}$ we have $\mathsf{En1}(\mathsf{U}) = \mathsf{U}\|\mathsf{Rd}(\mathsf{U})$, which justifies (22). Random variable $\mathsf{R} = \mathsf{Rd}(\mathsf{U})$ takes on at most $2^r$ values. Let $\mathsf{X} = \mathsf{U}$ and $\mathsf{Z} = \mathsf{Ch}_u(\mathsf{U})$ and apply Lemma 5.11 to get (23). ▮

## Acknowledgments

This work was supported by DARPA under contract HR011-09-C-0129. The views expressed are those of the authors and do not reflect the official policy or position of the Department of Defense or the U.S. Government.

# A Related Work

This section provides a comprehensive survey of the large body of work related to wiretap security, and more broadly, to information-theoretic secure communication in a noisy setup.

The wiretap setting and the secrecy capacity. For sake of clarity in comparing existing results, it is convenient to consider two different variants of the notion of secrecy capacity addressed by the wiretap literature. Given a pair of channels $(\mathsf{ChR}, \mathsf{ChA})$, their *weak secrecy capacity* $C_w$ is the supremum of the rates achievable by pairs $(\overline{\mathcal{E}}, \overline{\mathcal{D}})$ consisting of an encryption scheme $\overline{\mathcal{E}} = \{\mathcal{E}_k\}_{k\in\mathbb{N}}$ and a decryption scheme $\overline{\mathcal{D}} = \{\mathcal{D}_k\}_{k\in\mathbb{N}}$ for $\overline{\mathcal{E}}$ and $\mathsf{ChR}$ such that, first, the decoding requirement is satisfied, i.e.,

$$\lim_{k\to\infty} \mathbf{DE}(\mathcal{E}_k; \mathcal{D}_k; \mathsf{ChR}) = 0 \ .$$

and moreover,

$$\lim_{k\to\infty} \frac{\mathbf{Adv}^{\mathrm{res}}(\mathcal{E}_k; \mathsf{ChA})}{k} = 0 \ , \tag{24}$$

a property which we refer to as *weak security* in the following. Additionally, the *strong secrecy capacity* $C_s \leq C_w$ is obtained where we restrict the supremum over those schemes achieving RES security, i.e.,

$$\lim_{k\to\infty} \mathbf{Adv}^{\mathrm{res}}(\mathcal{E}_k; \mathsf{ChA}) = 0 \ , \tag{25}$$

It should be noted that both these quantities exhibit some drawbacks making them less appealing to cryptographers: First, the supermum is taken over schemes that are not necessarily computationally efficient. Moreover, both definitions only consider uniformly distributed messages. Finally, the privacy requirement of $C_w$, i.e., weak security, permits complete leakage of a $k^{1-\epsilon}$ plaintext bits, which is almost always unacceptable. In fact, cryptographic applications call for an even stronger capacity notion where $\mathbf{Adv}^{\mathrm{ds}}(\mathcal{E}_k; \mathsf{ChA})$ is a negligible function in $k$ (e.g., $2^{-\Theta(k)}$), and that this rate is achievable by an efficiently implementable scheme. Our main technical result can be interpreted as showing that such stringent capacity notion is not smaller than $C_w$ for many settings of interest, and it attained by an efficient and explicit scheme, which we provide.

Earlier work on the secrecy capacity. The wiretap scenario was first considered by Wyner [45], who provided a full characterization of $C_w$ in the special case where $\mathsf{ChA}$ is a degraded version of $\mathsf{ChR}$, i.e., such that there exists a transform $T$ with $T \circ \mathsf{ChR} = \mathsf{ChA}$, where the composition operator $\circ$ is the straightforward generalization of function composition to randomized transforms. Wyner's result was later generalized by Csiszár and Körner [13], who showed[2] that for arbitrary channels $\mathsf{ChR} : X \to Y$ and $\mathsf{ChA} : X \to Z$,

$$C_w = \max_{\mathsf{U},\mathsf{X}} \left[\mathbf{I}(\mathsf{U}; \mathsf{ChR}(\mathsf{X})) - \mathbf{I}(\mathsf{U}; \mathsf{ChA}(\mathsf{X}))\right] \ ,$$

where the maximum is over all pairs of correlated random variables $\mathsf{U}$ and $\mathsf{X}$, with the latter variable taking values from the channel domain. If both $\mathsf{ChR}$ and $\mathsf{ChA}$ are symmetric, the above simplifies to (cf. e.g. [26] for a proof)

$$C_w = \mathbf{H}(\mathsf{U}|\mathsf{ChA}(\mathsf{U})) - \mathbf{H}(\mathsf{U}|\mathsf{ChR}(\mathsf{U})) \tag{26}$$

where $\mathsf{U}$ is uniform on $X$. Therefore, in the most common setting of $\mathsf{ChR} = \mathsf{BSC}_{p_R}$ and $\mathsf{ChA} = \mathsf{BSC}_{p_A}$ (for $p_R < p_A$), (26) yields $C_w = h_2(p_A) - h_2(p_R)$. Finally, we note that more recently, Bloch and Laneman [6] have extended these results to the strong secrecy capacity $C_s$. Their work also considered other capacity measures related to security metrics different than those considered in this paper; however, all of these notions only cover random-input adversaries.

We stress that all aforementioned results are inherently *non-explicit*: That is, existence of secrecy-capacity achieving schemes is proven via the probabilistic method, and the resulting scheme is neither explicitly given, nor it is guaranteed to be efficient. In fact, to date, only a handful of efficient schemes are known.

[2] Csiszár and Körner actually considered a general setting where the outputs of $\mathsf{ChR}$ and $\mathsf{ChA}$ are correlated. However, we note that as long as communication is one-way, such correlation is irrelevant and one treats both (marginal) channels individually for correct decription and for security, respectively.

Wiretap channel II. Ozarow and Wyner [37] also considered an alternative to the above wiretap setting (called the *wiretap channel II*) where $\mathsf{ChR}$ is noiseless, but at the same time, Eve can learn a fraction $\delta$ of the bits sent over $\mathsf{ChR}$, and does not learn anything about the reamining $(1-\delta)$ fraction. Solutions were presented relying on error-correcting codes [37, 44]. Also, the notable work of [10] noted the such protocols with good parameters can be built from primitives such as deterministic randomness extractors for symbol-fixing sources with efficient inversion [23], as well as from $k$-wise independent functions [25] and related tools from exposure-resilient cryptography, such as all-or-nothing transforms [8, 17].

Information-theoretic key agreement. A technically and conceptually related line of work considers the setting of information-theoretically secure key agreement from noisy primitives where Alice and Bob can additionally *interact* via a noiseless *public* channel. The typical question is to study the *secret-key rate*, i.e., the maximal achievable ratio by a secret-key agreement scheme between the number of uses of the noisy primitive and the number key bits geneated by the protocol. (In general, there is no bound on the amount of communication over the public channel.) Note that key agreement and encryption are equivalent in this setting.

Maurer [30] introduced the problem and first showed that in Wyner's wiretap setting with $\mathsf{ChR} = \mathsf{BSC}_{p_R}$ and $\mathsf{ChA} = \mathsf{BSC}_{p_A}$, a higher secret-key rate $h_2(p_R + p_A - 2p_R p_A) - h_2(p_R)$ is achievable than in the pure wiretap setting. He later generalized [31] the question to a setting where Alice, Bob, and Eve are given each independent samples of random variables $\mathsf{X}, \mathsf{Y}$, and $\mathsf{Z}$, respectively. This seminal work was followed by several results following different directions, such as improving techniques used in these protocols [4, 7, 40], understanding the achievable secret-key rates in different settings [34, 39, 21] as well as extending to a setting where the channel between Alice and Bob is not authentic [33, 35, 24, 18, 9].

We also note that many of the tools developed in this area found interesting applications in the context of cryptography based on biometrics (starting from [15]).

Other wiretap models. We note that there is an active area of research aimed at understanding the secrecy and the secret-key capacities in even more settings, such as bi-directional wiretap settings (cf. e.g. [1]), multi-party settings, and channels with continous noise. We refer the reader e.g. to [5] for a survey of these results.

# B Proofs for Section 4

## B.1 Proof of Lemma 4.2

**Proof:** We have

$$\begin{aligned}
\mathbf{I}(\mathsf{M};\mathsf{C}) &= -\sum_{m} P_{\mathsf{M}}(m)\lg P_{\mathsf{M}}(m) + \sum_{c} P_{\mathsf{C}}(c)\sum_{m} P_{\mathsf{M}|\mathsf{C}=c}(m)\lg P_{\mathsf{M}|\mathsf{C}=c}(m)\\
&= -\sum_{m,c} J_{\mathsf{M},\mathsf{C}}(m,c)\lg P_{\mathsf{M}}(m) + \sum_{c} J_{\mathsf{M},\mathsf{C}}(m,c)\sum_{m}\lg P_{\mathsf{M}|\mathsf{C}=c}(m)\\
&= -\sum_{m,c} J_{\mathsf{M},\mathsf{C}}(m,c)\cdot\big(\lg P_{\mathsf{M}}(m) - \lg P_{\mathsf{M}|\mathsf{C}=c}(m)\big)\\
&= -\sum_{m,c} J_{\mathsf{M},\mathsf{C}}(m,c)\cdot\lg\frac{P_{\mathsf{M}}(m)}{P_{\mathsf{M}|\mathsf{C}=c}(m)}\\
&= \sum_{m,c} J_{\mathsf{M},\mathsf{C}}(m,c)\cdot\lg\frac{P_{\mathsf{M}|\mathsf{C}=c}(m)}{P_{\mathsf{M}}(m)}\\
&= \sum_{m,c} J_{\mathsf{M},\mathsf{C}}(m,c)\cdot\lg\frac{J_{\mathsf{M},\mathsf{C}}(m,c)}{P_{\mathsf{M}}(m)\cdot P_{\mathsf{C}}(c)}\\
&= \mathbf{D}(J_{\mathsf{M},\mathsf{C}};I_{\mathsf{M},\mathsf{C}})
\end{aligned}$$

as desired. ▌